\documentclass[aps,amsmath,twocolumn,amssymb,floatfix,showpacs,superscriptaddress,longbibliography]{revtex4-1}
\usepackage{amsmath}
\usepackage{mathtools}
\usepackage{braket}
\usepackage{bm}
\usepackage{graphicx}
\usepackage{amssymb}
\usepackage[dvipsnames]{xcolor}
\usepackage{multirow}
\usepackage{xfrac}
\usepackage{textcomp}
\usepackage{float}
\usepackage{enumerate}
\usepackage{subfigure}

\usepackage[colorlinks=true,linktoc=page,linkcolor=Plum,citecolor=blue,urlcolor=purple]{hyperref}

\newcommand{\vect}[1]{\bm{\mathrm{#1}}}

\newcommand{\ie}{{\it i.e.,\,\,}}
\newcommand{\eg}{{\it e.g.,~}}

\newcommand\bea{\begin{eqnarray}}
\newcommand\eea{\end{eqnarray}}
\newcommand\beq{\begin{equation}}  
\newcommand\eeq{\end{equation}}

\newcommand{\non}{\nonumber}  
\usepackage[normalem]{ulem}
\definecolor{lime}{HTML}{A6CE39}
\usepackage{sidecap,tikz}
\DeclareRobustCommand{\orcidicon}{\hspace{-1.0mm}
	\begin{tikzpicture}
	\draw[lime, fill=lime] (0.0,0.0) 
	circle [radius=0.15] 
	node[white] {{\fontfamily{qag}\selectfont \tiny \,ID}};
	\draw[white, fill=white] (-0.0525,0.095) 
	circle [radius=0.007];
	\end{tikzpicture}
	\hspace{-3.0mm}
}
\foreach \x in {A, ..., Z}{\expandafter\xdef\csname orcid\x\endcsname{\noexpand\href{https://orcid.org/\csname orcidauthor\x\endcsname}
		{\noexpand\orcidicon}}
}

\begin{document}
	

\title{Systematic generation of the cascade of anomalous dynamical first and higher-order modes in Floquet topological insulators}

\author{Arnob Kumar Ghosh\orcidA{}}
\email{arnob@iopb.res.in}
\affiliation{Institute of Physics, Sachivalaya Marg, Bhubaneswar-751005, India}
\affiliation{Homi Bhabha National Institute, Training School Complex, Anushakti Nagar, Mumbai 400094, India}
\author{Tanay Nag\orcidB{}}
\email{tnag@physik.rwth-aachen.de}
\affiliation{Institut f\"ur Theorie der Statistischen Physik, RWTH Aachen University, 52056 Aachen, Germany}
\author{Arijit Saha\orcidC{}}
\email{arijit@iopb.res.in}
\affiliation{Institute of Physics, Sachivalaya Marg, Bhubaneswar-751005, India}
\affiliation{Homi Bhabha National Institute, Training School Complex, Anushakti Nagar, Mumbai 400094, India}

\begin{abstract}
After extensive investigation on the Floquet second-order topological insulator (FSOTI) in two dimension (2D), here we propose two driving schemes to systematically engineer the hierarchy of Floquet first-order topological insulator, FSOTI, and Floquet third-order topological insulator in three dimension (3D). Our driving protocols allow these Floquet phases to showcase regular $0$, anomalous $\pi$, and hybrid $0$-$\pi$-modes in a unified phase diagram, obtained for both 2D and 3D systems, while staring from the lower order topological or non-topological phases. Both the step drive and the mass kick protocols exhibit the analogous structure of the evolution operator around the high symmetry points. These eventually enable us to understand the Floquet phase diagrams analytically and the Floquet higher order modes numerically based on finite size systems. The number of $0$ and $\pi$-modes can be tuned irrespective of the frequency in the step drive scheme while we observe frequency driven topological phase transitions for the mass kick protocol. We topologically characterize some of these higher order Floquet phases (harboring either $0$ or anomalous $\pi$ mode) by suitable topological invariant in 2D and 3D cases. 
\end{abstract}

\maketitle

\section{Introduction}

With the discovery of the quantum Hall effect~\cite{klitzing80}, the topological electronic materials have gained immense attention for their gapless boundary modes as caused by the non-trivial winding 
of the ground state wavefunction of the gapped bulk systems~\cite{thouless82}. Instead of the applying an external magnetic field, the quantum anomalous Hall effect then opens up a new avenue of research to be ventured in future~\cite{haldane88}. The introduction of spin-orbit coupling further enriches the research activity in this field when the time reversal symmetry (TRS) preserves 
quantum spin Hall insulator~\cite{kane05,moore07,roy09}, thus giving birth to the famous idea of topological insulator (TI)~\cite{BHZScience2006,hasan10,xlqi11}. 
Consequently, the concept of TI has been contextualized in several real materials~\cite{KonigScinece2007,zhang2009topological,hsieh2008topological} 
paving the way for the realization of bulk-boundary correspondence where the gapless boundary modes are the outcome of the topological band structure of the bulk crystal.
These can be exemplified for the first-order TI. Very recently, the concept of higher-order topological insulators (HOTIs)~\cite{benalcazar2017,benalcazarprb2017,Song2017,Langbehn2017,schindler2018,Franca2018,wang2018higher,Ezawakagome,Roy2019,Trifunovic2019,Khalaf2018,Szumniak2020,BiyeXie2021} have
received enormous attention in modern quantum condensed matter research. The bulk-boundary correspondence is generalized such that a $d$-dimensional $n^{\rm th}$ order HOTI is portrayed by the emergence of $(d-n)$-dimensional boundary modes. In particular, a two (three)-dimensional (2D (3D)) second-order topological insulator (SOTI) hosts zero (one) dimensional (0D (1D)) corner (hinge) modes, whereas a three-dimensional third-order topological insulator (TOTI) hosts 0D corner modes.

Moving our attention to the systems out of equilibrium, we emphasize that periodically driven quantum systems~\cite{Eckardt2017,oka2019} exhibit intriguing properties as compared to their static counterparts, such as dynamical localization~\cite{kayanuma08,nag14,nag15}, many-body localization~\cite{d13many,d14long,ponte15periodically}, Floquet time crystals~\cite{else16floquet,khemani16phase}, and higher harmonic generation~\cite{nag17,ikeda18} etc. Intriguingly, nondissipative dynamical boundary modes, owing to the time translational symmetry, can be contrived in Floquet topological insulator~\cite{oka09photovoltaic,kitagawa11transport,lindner11floquet,Rudner2013,nag2021anomalous,NHLindner2020} and Floquet topological superconductor~\cite{thakurathi13,benito14,sacramento15,rxzhang21}, which are the prime focus of interest in recent times. The nontrivial winding of the wavefunction for these driven systems in the time direction further allows one to have anomalous boundary modes namely, $\pi$-modes at finite energy. The prodigious experimental development of Floquet systems based on solid-state setup~\cite{Wang2013}, acoustic systems~\cite{peng2016experimental,fleury2016floquet}, photonic platforms~\cite{RechtsmanExperiment2013,MaczewskyExperiment2017} etc., add further merits to this field towards their realization and possible device application. Interestingly, the Floquet engineering also enables one to achieve Floquet HOTI (FHOTI) and HOT superconductor phases starting from a lower order or non-topological phase by suitably tuning appropriate driving protocols~\cite{Nag19,Bomantara2019,Seshadri2019,Martin2019,Huang2020,Ghosh2020,YangPRL2019,YangPRR2020,Nag2020,ZhangYang2020,HuPRL2020,GongPRBL2021,ghosh2020floquet,ghosh2020floquet2,
JiabinYu2021,Vu2021}.

Given the above background on the static and driven topological systems, we would like to emphasize that the question on the systematic generation of the anomalous HOTI phases 
(anchoring both regular $0$ and anomalous $\pi$-mode), in both 2D and 3D, has not been explored so far, to the best of our knowledge. Although, a few attempts have been made by activating the 
tight-binding piece-wisely in a step-like manner to show the emergence of 
anomalous $\pi$ corner modes in 2D~\cite{GongPRBL2021,HuPRL2020,Huang2020}. We note that the FHOTI states are not necessarily formulated from the underlying lower-order topological state. 
On the other hand, very recently, regular static zero-energy corner and gapless hinge modes are being shown in a hierarchical manner starting from the lower-order topological states~ \cite{Nag19,Nag2020,ghosh2021ladder}. Therefore, merging the above two aspects together, an immediate question arises that how to generate the flow of anomalous first-order topological insulator 
(FOTI) and HOTI phases starting from a trivial or static topological phase in both 2D and 3D. We refer to these phases as Floquet FOTI (FFOTI), Floquet SOTI (FSOTI), and Floquet TOTI (FTOTI).  
Ours is the first attempt to address these intriguing issues which have not been reported so far in the literature. 

In this article, we consider a step-like driving scheme, incorporating appropriate Wilson-Dirac mass terms in a simple tight-binding model to explore the appearance of the dynamical FFOTI as well as  
all the FHOTI phases in 2D and 3D. Interestingly, we find that the FFOTI phase diagrams, hosting $0$, $\pi$, $0$-$\pi$ edge (surface) modes (see Figs.~\ref{2Dfirst} and \ref{3Dfirst}), are directly 
upgraded into FSOTI (FSOTI and FTOTI) phase diagrams in 2D (3D) conceiving $0$, $\pi$, $0$-$\pi$ corner (hinge and corner) modes as shown in Fig.~\ref{2Dsecond} (Figs.~\ref{3Dsecond} and \ref{3Dthird}). Without restricting ourselves to the above particular driving scheme, we also exemplify the identical findings by considering mass kick protocol (see Fig.~\ref{Masskick}), where we find frequency-driven topological phase transitions in addition to the parameter-driven phase transitions. We analytically analyze the possible reason behind such systematic formulation of the FFOTI and the FHOTI phases from the Floquet operator that can signal the non-equilibrium phase diagram unanimously. We characterize some these phases by appropriate topological invariant such as, tangential polarization and octupolar moments (see Fig.~\ref{tanPol}). Furthermore, we illustrate the mosaic phase diagram, consisting of phases with different numbers of $0$ and $\pi$ modes, as a function of parameters associated with the driving (see Fig.~\ref{Extd}). We discuss several ancillary aspects of our analysis such as high-frequency limit, consequences of laser driving, effect of disorder etc., to highlight the importance of our findings. We believe that our work is experimentally viable provided the recent advancements in solid-state materials~\cite{schindler2018higher,Experiment3DHOTI.VanDerWaals}, mechanics~\cite{serra2018observation}, acoustics~\cite{xue2019acoustic}, microwaves~\cite{peterson2018quantized}, photonics~\cite{mittal2019photonic}, and electrical circuits~\cite{imhof2018topolectrical} etc. 

The remainder of the paper is structured as follows. In Sec.~\ref{Sec:II}, we introduce our two types of driving protocols along with the model Hamiltonians. Sec.~\ref{Sec:III} is devoted to the discussion of the main results of the paper, where we present the anomalous dynamical modes both in 2D and 3D for the FFOTI and the FHOTIs. In Sec.~\ref{Sec:IV}, we discuss the topological characterization of the FHOTI phases. After that, we discuss other possible approaches to engineer HOTI phases and their stabilization against disorder and a few related aspects in Sec.~\ref{Sec:V}. We finally summarize and conclude our paper in Sec.~\ref{Sec:VI}.

\vskip -1.0 cm
\section{Driving Protocols}\label{Sec:II}
Here we enlist our driving protocols in the form of step drive and periodic mass kick for both 2D and 3D.
\subsection{Step drive}
The step drive, consisting of two Hamiltonians that one can employ piece-wise within the time period to generate the FFOTI and the FHOTI phase. The driving scheme is explicitly demonstrated below
\begin{eqnarray}
H_{d \rm D}&=& J_1' h_{1, d \rm D}(\vect{k}) \  ,   \quad \quad t \in \Big[0,\frac{T}{2}\Big] \nonumber \\
&=&J_2' h_{2, d \rm D}(\vect{k}) \  , \quad \quad t \in \Big(\frac{T}{2},T \Big] \ ,
\label{drive1}
\end{eqnarray}
where, $J_1'$ and $J_2'$ carry the dimension of energy. We work with the natural unit, which allows us to set $\hbar=c=1$. The drive, however, is controlled by the dimensionless parameters $(J_1,J_2)=(J_1' T, J_2' T)$; where, $T$ being the period of the drive. This is related to the driving frequency $\Omega$ as $T=2 \pi / \Omega$. Here, $J'_i h_{i, d \rm D}(\vect{k})$ represents the  Hamiltonian of the system at the $i^{\rm th}$ step in $d$-dimension. In particular, to generate a 2D topological phase \cite{Nag19}, we consider $h_{1, 2 \rm D}(\vect{k})=\sigma_z $ and $h_{2, 2 \rm D}(\vect{k})=\left( \cos k_x + \cos k_y \right) \sigma_z+\sin k_x \sigma_x s_z + \sin k_y \sigma_y + \alpha \left( \cos k_x - \cos k_y \right)  \sigma_x s_x $; whereas, in 3D~\cite{Nag2020}, we envisage $h_{1, 3 \rm D}(\vect{k})=\mu _x \sigma_z $ and $h_{2, 3 \rm D}(\vect{k})=\left( \cos k_x + \cos k_y + \cos k_z \right) \mu _x \sigma_z+ \sin k_x \mu_x \sigma_x s_x  + \sin k_y\mu_x \sigma_x s_y + \sin k_z \mu_x \sigma_x s_z + \alpha \left( \cos k_x - \cos k_y \right)\mu_x \sigma_y + \beta \left( 2 \cos k_z -\cos k_x - \cos k_y \right) \mu_z$. Here, $\alpha$ and $\beta$ are dimensionless parameters, which we tune to generate the cascade of FFOTI, FSOTI, and FTOTI phases. The three Pauli matrices $\bm \mu$, $\bm \sigma$, and $\bm s$ act on sublattice $(A,B)$, orbital $(a,b)$, and spin $(\uparrow,\downarrow)$ degrees of freedom, respectively.

Note that the first (second) step Hamiltonian $J'_1 h_{1, d \rm D}(\vect{k})$ ($J'_2 h_{2, d \rm D}(\vect{k})$) is composed of on-site (hopping) terms only. We sometime refer to $h_{1, d \rm D}(\vect{k})$ 
as $h_{1, d \rm D}$  due to its on-site nature. The $\alpha$ and $\beta$ independent cosine terms $\cos k_{x,y,z}$ arise due to the nearest-neighbor hopping while the spin orbit coupling is represented by the sine terms $\sin k_{x,y,z}$. The on-site mass term of strength $J'_1$ becomes very important for topological phase transition that we discuss below. The first step Hamiltonian, thus preserves the necessary symmetries while the second step Hamiltonian is found to be responsible for breaking certain symmetries. The latter becomes very much useful to achieve the FHOTI phases. For $\alpha=\beta=0$, $h_{i, d \rm D}(\vect{k})$ respects TRS generated by $\mathcal{T}=i s_y \mathcal{K}$; $\mathcal{K}$ being the complex-conjugation operator. Although when $\alpha \neq 0$, the corresponding term, thereby $h_{2, d \rm D}(\vect{k})$ breaks both TRS and four-fold rotation ($C_4$) symmetry while preserving the combined $C_4 \mathcal{T}$ symmetry.  Importantly,  $h_{i, d \rm D}(\vect{k})$ respects unitary chiral ${\mathcal C}= \sigma_x s_y$ ($=\mu_y \sigma_0 s_0$) and anti-unitary particle-hole symmetry ${\mathcal P}=\sigma_x s_z \mathcal{K}$ ($=\mu_x \sigma_y s_y \mathcal{K}$) for 2D (3D) such that ${\mathcal C} h_{i, d \rm D}(\vect{k}) {\mathcal C}^{-1}= -h_{i, d \rm D}(\vect{k})$ and ${\mathcal P} h_{i, d \rm D}(\vect{k}) {\mathcal P}^{-1}= -h_{i, d \rm D}(-\vect{k})$~\cite{Nag19,Nag2020}.
It is to be noted that the term associated with $\alpha$ further breaks the mirror symmetry.

Before translating to the dynamical limit, we would like to point out the various static phases accessible to our model, depending upon the values of different parameters of the system. 
One can contemplate the following Hamiltonians in 2D and 3D as
\begin{eqnarray}
H_{2 \rm D}^{\rm Static}(\vect{k})&=& J_1' h_{1, 2 \rm D}(\vect{k}) + J_2' h_{2, 2 \rm D}(\vect{k}) \ , \label{2DStatic} \\
H_{3 \rm D}^{\rm Static}(\vect{k})&=& J_1' h_{1, 3 \rm D}(\vect{k}) + J_2' h_{2, 3 \rm D}(\vect{k}) \ , \label{3DStatic}
\end{eqnarray}
where, $H_{2 \rm D}^{\rm Static}(\vect{k})$ represents the Hamiltonian of a 2D quantum spin Hall insulator (QSHI) with propagating 1D helical edge states, when $\alpha=0$ and $0< \lvert J_1' \rvert < 2 \lvert J_2' \rvert $~\cite{BHZScience2006,KonigScinece2007}. However, for any non-zero value of $\alpha$, the edge states of the QSHI are gapped out by the corresponding Wilson-Dirac mass term proportional to $\alpha$ in such a way that two consecutive edges incorporate opposite mass terms. Using the generalized Jackiw-Rebbi index theorem~\cite{JWPRD1976}, 
one can show the emergence of zero-modes at the corners of the system \ie the system resides in the SOTI phase~\cite{schindler2018,Nag19}.

In 3D, we have access to an additional HOT phase namely, TOTI for $\alpha,~\beta\ne 0$ as compared to the 2D case with $\alpha\ne 0$ only, where one can reach up to the SOTI. We here point out this hierarchy one by one. In absence of the Wilson-Dirac masses \ie $\alpha=\beta=0$, $H_{3 \rm D}^{\rm Static}(\vect{k})$ hosts gapless 2D surface states in the strong TI phase provided $0< \lvert J_1' \rvert < 3 \lvert J_2' \rvert $,~\cite{zhang2009topological,Nag2020,ghosh2021ladder}. On top of the 3D TI phase, if we allow $\alpha$ to be non-zero, the surface states of the underlying TI are gapped out 
by the corresponding term, and we procure gapless states along the 1D hinge of the system in the $z$-direction~\cite{benalcazar2017,schindler2018,Nag2020,ghosh2021ladder}. This is the signature of a 3D SOTI. On the other hand, in presence of non-zero values of $\alpha$ and $\beta$, the first order 3D TI phase transmutes into a 3D TOTI, harboring 0D corner modes~\cite{benalcazar2017,Nag2020,ghosh2021ladder}. 
 
 \vspace {-0.2cm}

\subsection{Periodic mass kick}
In our periodic kick protocol, we consider the Hamiltonian $\mathcal{H}_{d \rm D}$ in $d$-dimension between two successive kicks. Afterwards, we introduce the driving protocol in the form 
of an on-site mass kick as 
\begin{eqnarray} 
m_0(t)&=& m \ h_{1, d \rm D} \sum_{r=1}^{\infty} \delta(t-rT) \ , 
\label{kick1}
\end{eqnarray}
where, $m$ denotes the strength of the kicking parameter, $t$ denotes time, and $T$ symbolizes the time-period of the drive. Following the periodic kick, we can write down the exact Floquet operator, $U_{d \rm D}(\vect{k},T)$ using the time-ordered ($\overline{\rm TO}$) notation as
\begin{eqnarray}
U_{d \rm D}(\vect{k},T)&=&\overline{\rm TO} \exp \left[-i\int_{0}^{T}dt\left(\mathcal{H}_{d \rm D}({\vect{k}})+m_0(t) \right)\right] \nonumber \\
&=& \exp(-i \mathcal{H}_{d \rm D}({\vect{k}}) T)~\exp(-i m~h_{1, d \rm D})\ .
\label{fomasskick}
\end{eqnarray}

We choose $\mathcal{H}_{2 \rm D}({\vect{k}})=J^{\prime} h_{2, 2 \rm D}(\vect{k})$ in 2D and $\mathcal{H}_{3 \rm D}({\vect{k}})=J' h_{2, 3 \rm D}(\vect{k})$ in 3D. Also, $h_{1, 2 \rm D}$ $h_{1, 3 \rm D}$
are defined earlier. Here, $J'$ carries the dimension of energy. However we use the dimensionless parameter $J=J'T$ along with $m$ to control the drive. Similar to Eqs.(\ref{2DStatic}) and (\ref{3DStatic}), one can obtain the analogous static Hamiltonian for this case as $H_{d \rm D}^{\rm Static}(\vect{k})= m \ h_{1, d \rm D} + J' h_{2, d \rm D}(\vect{k})$. Note that, one can achieve the 
mass kick (Eq.(\ref{kick1})) protocol from the limiting case of step drive protocol (Eq.(\ref{drive1})) with infinitesimal duration of the first step Hamiltonian $J'_1 h_{1, d \rm D}$.

\vspace {0.4cm}

\section{Anomalous dynamical modes}\label{Sec:III}
Here we present our key findings regarding the generation of dynamical FHOTI phases in 2D and 3D, employing the step drive and periodic mass kick protocols.
\vspace {0.1cm}
\subsection{Step drive}
Within the step drive protocol, we can generate the jets of FFOTI, FSOTI, and FTOTI by tuning some specific parameters, which we discuss in the upcoming subsections in detail.
\vspace {-0.2cm}
\subsubsection*{\rm {\bf{1.~~2D}}}
To begin with 2D, following step drive protocol, we can write down the full Floquet evolution operator $U_{2 \rm D}(\vect{k},T)$ after one full period $T$ as
\begin{eqnarray}\label{U2Dstep}
U_{2 \rm D}(\vect{k},T)= \exp\left(-i \frac{J_2}{2} h_{2, 2 \rm D}(\vect{k}) \right) \exp\left(-i \frac{J_1}{2} h_{1, 2 \rm D}(\vect{k})  \right)\ , \qquad
\end{eqnarray}
where, we can express $U_{2 \rm D}(\vect{k},T)$ as $U_{2 \rm D}(\vect{k},T)=f_{2 \rm D}(\vect{k})  \mathbb{I}+i g_{2 \rm D}(\vect{k})$, such that 
\begin{widetext}
\begin{eqnarray}
f_{2 \rm D}(\vect{k})&=& \cos\left(\gamma_{2 \rm D}(\vect{k})  \frac{J_1}{2}\right) \cos\left( \lambda_{2 \rm D}(\vect{k}) \frac{J_2}{2} \right) - \sin\left(\gamma_{2 \rm D}(\vect{k}) \frac{J_1}{2}\right) \sin\left( \lambda_{2 \rm D}(\vect{k}) \frac{J_2}{2} \right)  \chi_{2 \rm D}(\vect{k}) \ , \label{fk2Dstep}  \\
g_{2 \rm D}(\vect{k})&=& - \frac{1}{\gamma_{2 \rm D}(\vect{k}) \lambda_{2 \rm D}(\vect{k})}  \sin\left(\gamma_{2 \rm D}(\vect{k})  \frac{J_1}{2}\right) \sin\left( \lambda_{2 \rm D}(\vect{k})  \frac{J_2}{2} \right) \eta_{2 \rm D}(\vect{k}) -  \sin\left(\gamma_{2 \rm D}(\vect{k})  \frac{J_1}{2}\right) \cos \left( \lambda_{2 \rm D}(\vect{k})  \frac{J_2}{2} \right) \frac{h_{1, 2 \rm D}(\vect{k})}{\gamma_{2 \rm D}(\vect{k})} \non \\
&&-  \cos\left(\gamma_{2 \rm D}(\vect{k})  \frac{J_1}{2}\right) \sin \left( \lambda_{2 \rm D}(\vect{k})  \frac{J_2}{2} \right) \frac{h_{2, 2 \rm D}(\vect{k})}{\lambda_{2 \rm D}(\vect{k})} \  \label{gk2Dstep} \ .
\end{eqnarray}
\vskip -0.2cm
\end{widetext}

Here, we have suppressed the implicit $T$ dependence in $f_{2 \rm D}(\vect{k})$ and $g_{2 \rm D}(\vect{k})$. We have defined $\gamma_{2 \rm D}(\vect{k})=\lvert h_{1, 2 \rm D}(\vect{k}) \rvert$, $\lambda_{2 \rm D}(\vect{k})=\lvert h_{2, 2 \rm D}(\vect{k}) \rvert$, $\chi_{2 \rm D}(\vect{k})=\frac{\cos k_x +\cos k_y}{\gamma_{2 \rm D}(\vect{k}) \lambda_{2 \rm D}(\vect{k})}$, and $\eta_{2 \rm D}(\vect{k})= \sin k_x \sigma_y s_z - \sin k_y \sigma_x + \alpha \left( \cos k_x - \cos k_y \right) \sigma_y s_x $. From the eigenvalue equation for $U_{2 \rm D}(\vect{k},T)$: $U_{2 \rm D}(\vect{k},T) \ket{\Psi} = e^{i E(\vect{k})} \ket{\Psi}$, one obtains the condition
\begin{eqnarray}
\cos E(\vect{k})=f_{2 \rm D}(\vect{k}) \ ,
\end{eqnarray}
with each band being two-fold degenerate. Now, the band gap closes at $ \vect{k}=\vect{k}^*=(0,0)~{\rm or}~(\pi,\pi)$, for $f_{2 \rm D}(\vect{k}^*)=\pm 1$ such that $E(\vect{k}^*)=n\pi$ with $n=0,1,2,3,\cdots$ being an integer. The interesting point to note here is that we can cast $f_{2 \rm D}(\vect{k})$ in terms of a single cosine function such as $f_{2 \rm D}(\vect{k}^*)=\cos(\gamma_{2 \rm D}(\vect{k}^*)  \frac{J_1}{2} \pm \lambda_{2 \rm D}(\vect{k}^*) \frac{J_2}{2} )$. Furthermore, the structure of $\chi_{2 \rm D}(\vect{k}^*)$ term here serves as an essential ingredient to continue with the above analysis.  To be precise, at these momentum points $\gamma_{2 \rm D}(\vect{k}^*)$, $\lambda_{2 \rm D}(\vect{k}^*)$, and $\chi_{2 \rm D}(\vect{k}^*)$ take the values as follows $\gamma_{2 \rm D}(0,0)=\gamma_{2 \rm D}(\pi,\pi)=1$, $\lambda_{2 \rm D}(0,0)=\lambda_{2 \rm D}(\pi,\pi)=2$, and $\chi_{2 \rm D}(0,0)=-\chi_{2 \rm D}(\pi,\pi)=1$. These special momentum modes continue playing pivotal role for dynamics in addition to the static counter part as given in  
Eqs.(\ref{2DStatic}) and (\ref{3DStatic}). This enables us to write the right hand side~(RHS) 
of Eq.(\ref{fk2Dstep}) in a compact form as 
\begin{equation}\label{cosphase2D}
\cos \left( \frac{J_1}{2} \pm J_2  \right) = \cos n \pi \ ,
\end{equation}
where, $n$ is an integer. From Eq.~(\ref{cosphase2D}), we obtain the gap closing conditions in terms of the dimensionless parameters $J_1$ and $J_2$ as
\begin{equation}\label{phasestep2D}
\lvert J_2 \rvert  = \frac{\lvert J_1 \rvert}{2} + n \pi \ .
\end{equation}

\begin{figure}[]
	\centering
	\subfigure{\includegraphics[width=0.5\textwidth]{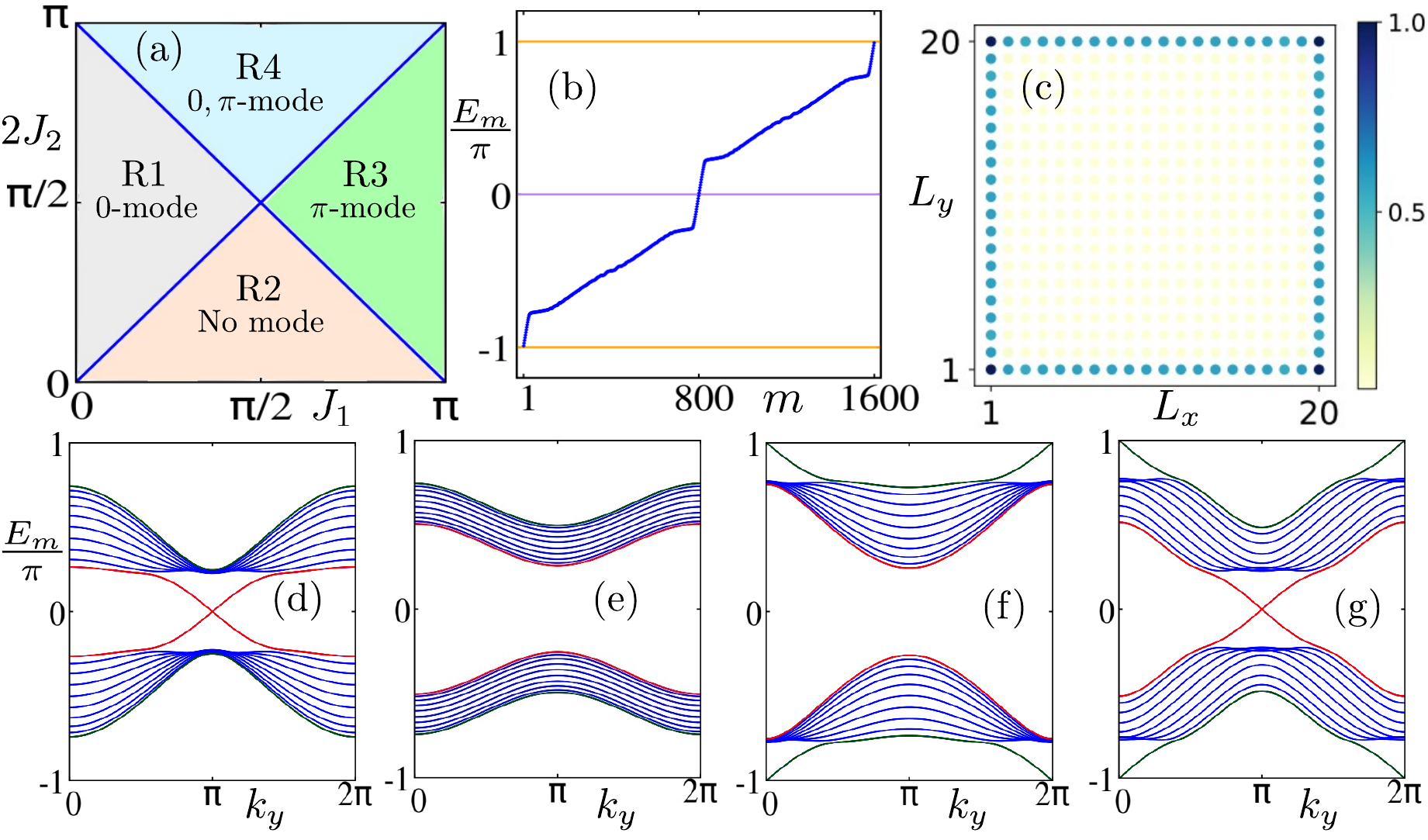}}
	\caption{(a) (Color online) Phase diagram is depicted in the $J_1-J_2$ plane for 2D FFOTI, 
which is guided by Eq.(\ref{phasestep2D}). (b) Quasi-energy spectrum for a finite system and as a function of the state index $m$, is depicted when the system is in the phase R4. One can clearly identify modes close to quasi-energies $0$ and $\pm \pi$. (c) The LDOS is shown for a finite-size system corresponding to quasi-energies $E_m=0,~\pm \pi$ in R4. The quasi-energy spectra $E_m$, considering slab geometry, for R1, R2, R3, and R4 are shown in panels (d), (e), (f), and (g), respectively. We choose the parameters as $(J_1,2 J_2)=\left[ \left(\frac{\pi}{4},\frac{\pi}{2}\right),~\left(\frac{\pi}{2},\frac{\pi}{4}\right),~\left(\frac{3\pi}{4},\frac{\pi}{2}\right), ~\left(\frac{\pi}{2},\frac{3 \pi}{4}\right) \right] $ for R1, R2, R3, and R4, respectively.
	}
	\label{2Dfirst}
\end{figure}

Here, Eq.~(\ref{phasestep2D}) signifies the topological phase boundaries between various dynamic phases, as depicted in Figs.~\ref{2Dfirst}(a) and ~\ref{2Dsecond}(a). The phase boundaries divide 
the phase diagram into four parts- region 1~(R1) with $0$-mode, region 2~(R2) with no modes, region 3~(R3) with $\pi$-mode, and region 4~(R4) hosting both $0$ and $\pi$-modes. Interestingly, the phase boundaries remain the same for both FFOTI and FSOTI due to the absence of any $\alpha$ dependent term in Eq.~(\ref{phasestep2D}). Instead of considering two step drive, one can also think about a three step drive protocol as discussed in Ref.~\cite{Huang2020} and end up obtaining a relation akin to Eq.~(\ref{phasestep2D}). The underlying reason for this can be attributed to the fact that the structure of $f_{2 \rm D}(\vect{k}^*)$ remains unaltered in both cases. With the help of Eqs.~(\ref{U2Dstep}), (\ref{fk2Dstep}), and (\ref{gk2Dstep}), we can write the Floquet effective Hamiltonian 
$H_{2 \rm D, Flq}$ as 
\begin{equation}\label{EffHam2Dstep}
 H_{2 \rm D, Flq}=- \frac{\epsilon_{2\rm D}(\vect{k})}{\sin \left[\epsilon_{2\rm D}(\vect{k}) T\right]} g_{2\rm D}(\vect{k}) \ ,
\end{equation}
where, $\epsilon_{2\rm D}(\vect{k})= \frac{1}{T}\cos^{-1}\left[f_{2\rm D}(\vect{k})\right]$.

Note that the ladder of FHOTI including FFOTI can be engineered by selectively incorporating the Wilson-Dirac masses in $h_{2, 2 \rm D}(\vect{k})$. To be precise, $\alpha=0$ ($\alpha \ne 0$) in $h_{2, 2 \rm D}(\vect{k})$ leads to FFOTI (FSOTI) phases. After analytical understanding of the emergence of FFOTI and FSOTI phases, we further support our findings by numerical analysis for finite size systems (see text for discussion).  
\vspace{-0.3cm}
\subsubsection*{\rm Case 1: FFOTI}
As mentioned earlier, in order to obtain the FFOTI phase, we set $\alpha=0$. The FFOTI is characterized by the presence of gapless edge modes at its 1D boundary. To uncover the subsistence of gapless \textit{dispersive} edge modes, we resort to slab geometry \ie periodic boundary condition~(PBC) along one direction (say $y$) and open boundary condition~(OBC) in the other direction~(say $x$). We numerically diagonalize the Floquet operator $U_{2 \rm D}(k_y,T)$ and depict the quasi-energy spectrum as a function of $k_y$ for regions R1, R2, R3, and R4 in Figs.~\ref{2Dfirst}(d), (e), (f), and (g), respectively. It is evident that the gapless modes appear at $k_y=0,~\pi$ as discussed in the mathematical analysis before. With reference to the phase diagram, Fig.~\ref{2Dfirst}(a), obtained analytically from Eq.(\ref{phasestep2D}) and verified numerically from $U_{2 \rm D}(k_y,T)$ (Eq.(\ref{U2Dstep})), one can observe only $0$-mode, no mode, only $\pi$-mode and both $0$ and $\pi$-mode in R1, R2, R3, and R4, respectively. The real-space quasi-energy spectrum associated with R4 is depicted in Fig.~\ref{2Dfirst}(b). There one can identify the FFOTI modes at both $E_m=0$ and $E_m=\pm\pi$.
As dicussed above, one expect to notice FFOTI modes separately at $E_m=0$ ($E_m=\pm \pi$) for R1 (R3) and no modes either at $E_m=0$ or $E_m=\pi$ for R2. Although we prefer not to show them here. The local density of states (LDOS) for a system with OBC in both directions is illustrated in Fig.~\ref{2Dfirst}(c), for quasi-states with $E_m=0,~\pm \pi$ (within numerical accuracy) corresponding to R4. 
The LDOS remains unaltered for quasi-states corresponding to $E_m=0$ (R1) and $E_m=\pm \pi$ (R3). These $E_{m}=\pm \pi~(\rm at~frequency~\pm \Omega/2)$ gapless modes refer to the dynamical ones as reported earlier for other systems with different driving scheme~\cite{Piskunow2014,Usaj2014}. 

\subsubsection*{\rm Case 2: FSOTI}
\begin{figure}[]
	\centering
	\subfigure{\includegraphics[width=0.48\textwidth]{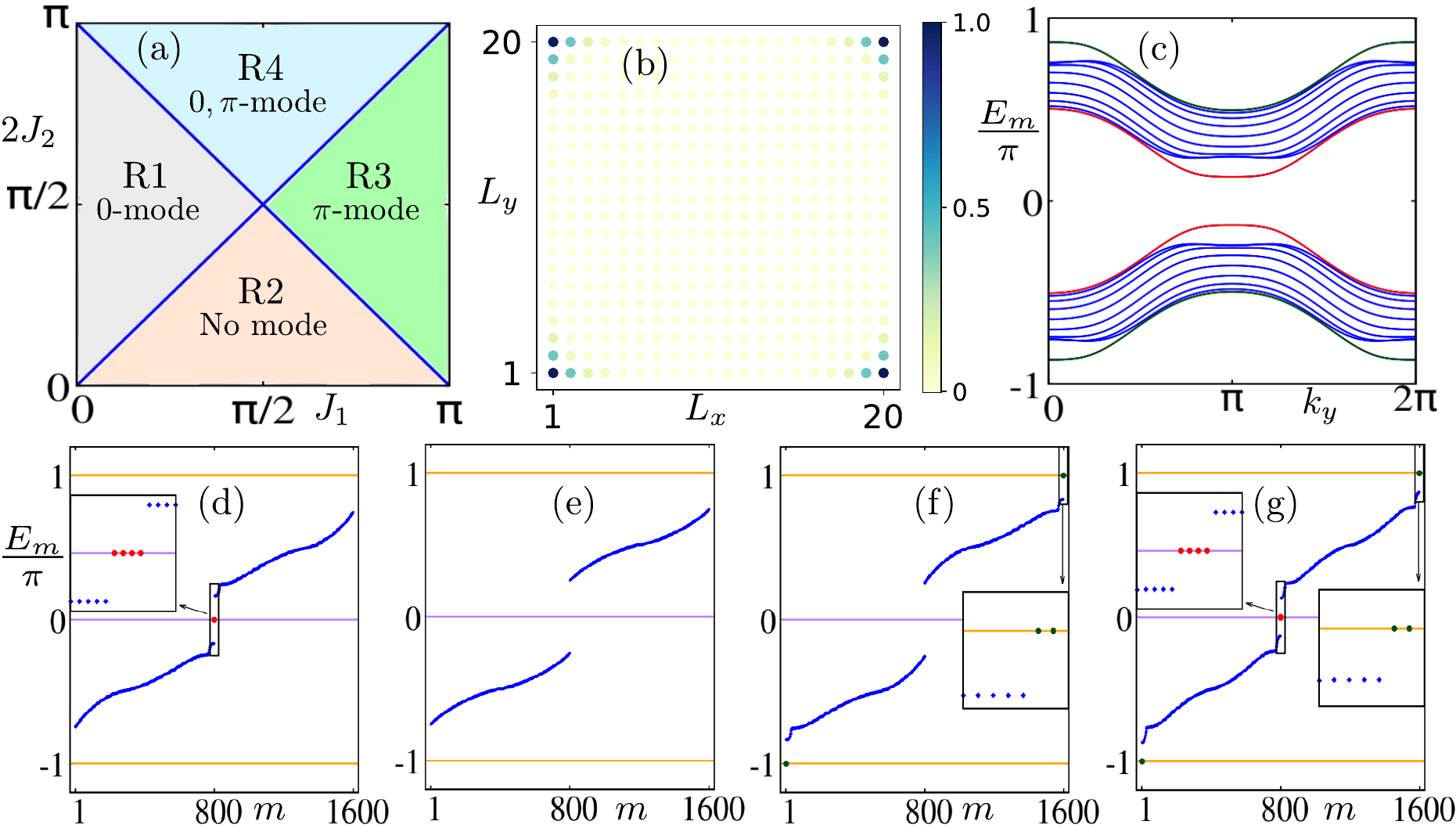}}
	\caption{(Color online)
	(a) Phase diagram is demonstrated in the $J_1-J_2$ plane for 2D FSOTI. (b) LDOS is shown 
for a finite-size system corresponding to quasi-energies $E_m=0,~\pm \pi$ in R4. (c) Quasi-energy spectrum $E_m$, considering slab geometry for our system, is depicted as a function of $k_y$ 
for R4. One can observe that both the $0$ and $\pi$ edge modes are gapped out due to inclusion of the Wilson-Dirac mass term $\alpha$. The quasi-energy spectra $E_m$ for the finite-size system, corresponding to R1, R2, R3, and R4 are presented as a function of the state index $m$ in panels (d), (e), (f), and (g), respectively. We use the same parameters as mentioned in  
Fig.~\ref{2Dfirst}.
	}
	\label{2Dsecond}
\end{figure}

Turning our focus to 2D FSOTI phase that can be obtained by considering $\alpha\ne 0$, the corresponding phase diagram comes out to be unaltered as depicted in Fig.~\ref{2Dsecond}(a).
To probe the footprints of FSOTI via the LDOS, we resort to OBC in both directions. We show the LDOS for quasi-states with $E_m=0,~\pm \pi$ in Fig.~\ref{2Dsecond}(b). It is evident that corner localized modes appear at quasi-energies $E_m=0,~\pm \pi$. The quasi-energy spectra are shown in Fig.~\ref{2Dsecond}(d), (e), (f), and (g) for the phase R1, R2, R3, and R4, respectively. As mentioned earlier, when we set $\alpha$ to be non-zero, the 1D edge-modes of FFOTI should be gapped out by the Wilson-Dirac mass term $(\cos k_x- \cos k_y)$ and the corresponding $0, \pi$ modes appear at the corners of the system. This is clearly visible for the present case while comparing Fig.~\ref{2Dfirst}(g) with Fig.~\ref{2Dsecond}(c), where we depict the gapped edge-mode in R4 employing slab geometry.  Note that, there exist $4~(2)$ quasi-states at quasi-energy $E_m=0,~(\pm \pi)$ for the 2D FSOTI phase.
\vskip +0.3cm

\subsubsection*{\rm {\bf{2.~~3D}}}
In 3D, we comprehensively unveil, how to systematically generate the cascade of FFOTI, FSOTI, and FTOTI phase, by suitably tuning the Wilson-Dirac mass perturbations. In 3D, the Floquet evolution operator for the step drive protocol reads as
\begin{eqnarray} \label{U3Dstep}
U_{3 \rm D}(\vect{k},T)=\exp\left(-i \frac{J_2}{2} h_{2, 3 \rm D}(\vect{k}) \right) \exp\left(-i \frac{J_1}{2} h_{1, 3 \rm D}(\vect{k})  \right) . \qquad
\end{eqnarray}
Similar to the 2D case, one can find the Floquet evolution operator $U_{3 \rm D}(\vect{k},T)$ as $U_{3 \rm D}(\vect{k},T)=f_{3 \rm D}(\vect{k})  \mathbb{I}+i g_{3 \rm D}(\vect{k})$, with 
\vskip -0.2cm
\begin{widetext}
	\begin{eqnarray}
	f_{3 \rm D}(\vect{k})&=& \cos\left(\gamma_{3 \rm D}(\vect{k})  \frac{J_1}{2}\right) \cos\left( \lambda_{3 \rm D}(\vect{k}) \frac{J_2}{2} \right) - \sin\left(\gamma_{3 \rm D}(\vect{k}) \frac{J_1}{2}\right) \sin\left( \lambda_{3 \rm D}(\vect{k}) \frac{J_2}{2} \right)  \chi_{3 \rm D}(\vect{k}) \ , \label{fk3Dstep}  \\
	g_{3 \rm D}(\vect{k})&=& - \frac{1}{\gamma_{3 \rm D}(\vect{k}) \lambda_{3 \rm D}(\vect{k})} \eta_{3 \rm D}(\vect{k})  \sin\left(\gamma_{3 \rm D}(\vect{k})  \frac{J_1}{2}\right) \sin\left( \lambda_{3 \rm D}(\vect{k})  \frac{J_2}{2} \right)-  \sin\left(\gamma_{3 \rm D}(\vect{k})  \frac{J_1}{2}\right) \cos \left( \lambda_{3 \rm D}(\vect{k})  \frac{J_2}{2} \right) \frac{h_{1, 3 \rm D}(\vect{k})}{\gamma_{3 \rm D}(\vect{k})} \non \\
	&&-  \cos\left(\gamma_{3 \rm D}(\vect{k})  \frac{J_1}{2}\right) \sin \left( \lambda_{3 \rm D}(\vect{k})  \frac{J_2}{2} \right) \frac{h_{2, 3 \rm D}(\vect{k})}{\lambda_{3 \rm D}(\vect{k})} \label{gk3Dstep} \ ,
	\end{eqnarray}
\end{widetext}
where, $\gamma_{3 \rm D}(\vect{k})=\lvert h_{1, 3 \rm D}(\vect{k}) \rvert$, $\lambda_{3 \rm D}(\vect{k})=\lvert h_{2, 3 \rm D}(\vect{k}) \rvert$, $\chi_{3 \rm D}(\vect{k})=\frac{\cos k_x +\cos k_y+\cos k_z}{\gamma_{3 \rm D}(\vect{k}) \lambda_{3 \rm D}(\vect{k})}$, and $\eta_{3 \rm D}(\vect{k})= \sin k_x \sigma_y s_x + \sin k_y \sigma_y s_y + \sin k_z \sigma_y s_z - \alpha \left( \cos k_x - \cos k_y \right) \sigma_x - \beta \left( 2 \cos k_z - \cos k_x - \cos k_y \right) \mu_y \sigma_z$. From the eigenvalue equation of $U_{3 \rm D}(\vect{k},T)$: $U_{3 \rm D}(\vect{k},T) \ket{\Psi} = e^{i E(\vect{k})} \ket{\Psi}$, one obtains
\vspace{-0.2cm}
\begin{eqnarray}
\cos E(\vect{k})=f_{3 \rm D}(\vect{k}) \ ,
\end{eqnarray}
\vspace{+0.3cm}
\noindent
with each band being four-fold degenerate. Now, the band gap closes at $\vect{k}=\vect{k}^*=(0,0,0)~{\rm or}~(\pi,\pi,\pi)$, when $f_{3 \rm D}(\vect{k})=\pm 1$. At these points $\gamma_{3 \rm D}(\vect{k})$, $\lambda_{3 \rm D}(\vect{k})$, and $\chi_{3 \rm D}(\vect{k})$ take the values as follows $\gamma(0,0,0)=\gamma(\pi,\pi,\pi)=1$, $\lambda(0,0,0)=\lambda(\pi,\pi,\pi)=3$, and $\chi(0,0,0)=-\chi(\pi,\pi,\pi)=1$. This enables us to write the right hand side of Eq.~(\ref{fk3Dstep}), following the similar line of arguments discussed for 2D case, as 
\begin{equation}
\cos \left( \frac{J_1}{2} \pm \frac{3 J_2}{2}  \right) = \cos n \pi \ ,
\label{3Dcond}
\end{equation}
From the above relation (Eq.~(\ref{3Dcond})), we obtain the gap closing relations in terms of $J_1$ and $J_2$ as
\begin{equation}\label{phasestep3D}
 \frac{3 \lvert J_2 \rvert}{2}   = \frac{\lvert J_1 \rvert}{2} + n \pi \ ,
\end{equation}
where, $n$ is an integer. Note that, Eq.~(\ref{phasestep3D}) serves the purpose of topological phase boundary in case of 3D. It is worth mentioning here that the phase diagrams, in the $J_1-J_2$ plane to obtain different type of first and higher order modes, remain the same for FFOTI, FSOTI, and FTOTI~(see Figs.~\ref{3Dfirst}(a), \ref{3Dsecond}(a), and \ref{3Dthird}(a), respectively). This can be attributed to the fact that gap-closing conditions are independent of $\alpha$ and $\beta$. This is a generic feature of this particular driving protocol. However, the special structures of $\chi_{3 \rm D}(\vect{k}^*)$ 
and $f_{3 \rm D}(\vect{k}^*)$ are held responsible for the above robust nature of these phase diagrams. Although, depending upon the choice of the driving scheme, one can obtain FHOTI phase diagram 
as a function of mass terms $\alpha$ and $\beta$. Using Eqs.(\ref{U3Dstep}), (\ref{fk3Dstep}), and (\ref{gk3Dstep}), we can write the Floquet effective Hamiltonian 
$H_{3 \rm D, Flq}$ as 
\begin{equation}\label{EffHam3Dstep}
H_{3 \rm D, Flq}=- \frac{\epsilon_{3\rm D}(\vect{k})}{\sin \left[\epsilon_{3\rm D}(\vect{k}) T\right]} g_{3\rm D}(\vect{k}) \ ,
\end{equation}
where, $\epsilon_{3\rm D}(\vect{k})= \frac{1}{T}\cos^{-1}\left[f_{3\rm D}(\vect{k})\right] $.

Similar to the earlier case for 2D system, we below systematically explore the emergence of FFOTI and FHOTI phases by numerically diagonalizing the Floquet operator (Eq.(\ref{U3Dstep}))  
implementing appropriate finite geometries.
\subsubsection*{\rm Case 1: FFOTI}
\begin{figure}[]
	\centering
	\subfigure{\includegraphics[width=0.48\textwidth]{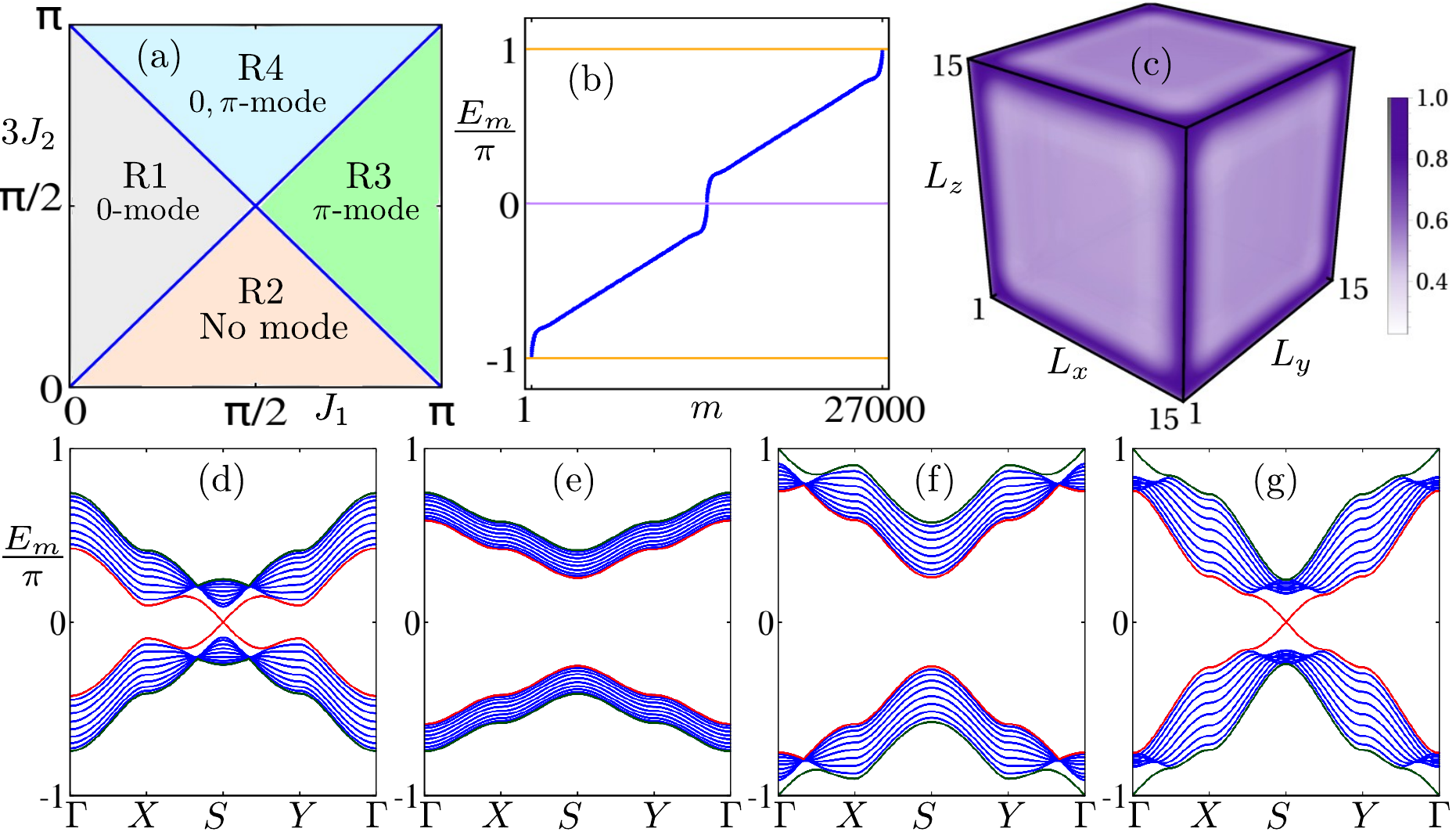}}
	\caption{(Color online)
	(a) Phase diagram is depicted in the $J_1-J_2$ plane for 3D FFOTI. (b) Quasi-energy spectrum, $E_m$ for a finite size system is shown as a function of the state index $m$, when the system is in  
	phase R4. (c) LDOS is depicted for finite-size system corresponding to quasi-energies $E_m=0,~\pm \pi$ for R4. The quasi-energy spectra $E_m$ cosidering slab geometry, along the 
	$\Gamma-X-S-Y-\Gamma$ points, for R1, R2, R3, and R4 are shown in panels (d), (e), (f), and (g), respectively. The chosen parameters are: $(J_1,3 J_2)=\left[ \left(\frac{\pi}{4},\frac{\pi}{2}\right),~
	\left(\frac{\pi}{2},\frac{\pi}{4}\right),~\left(\frac{3\pi}{4},\frac{\pi}{2}\right),~\left(\frac{\pi}{2},\frac{3 \pi}{4}\right) \right]$ for R1, R2, R3, and R4, respectively.	
	}
	\label{3Dfirst}
\end{figure}

\begin{figure}[]
	\centering
	\subfigure{\includegraphics[width=0.48\textwidth]{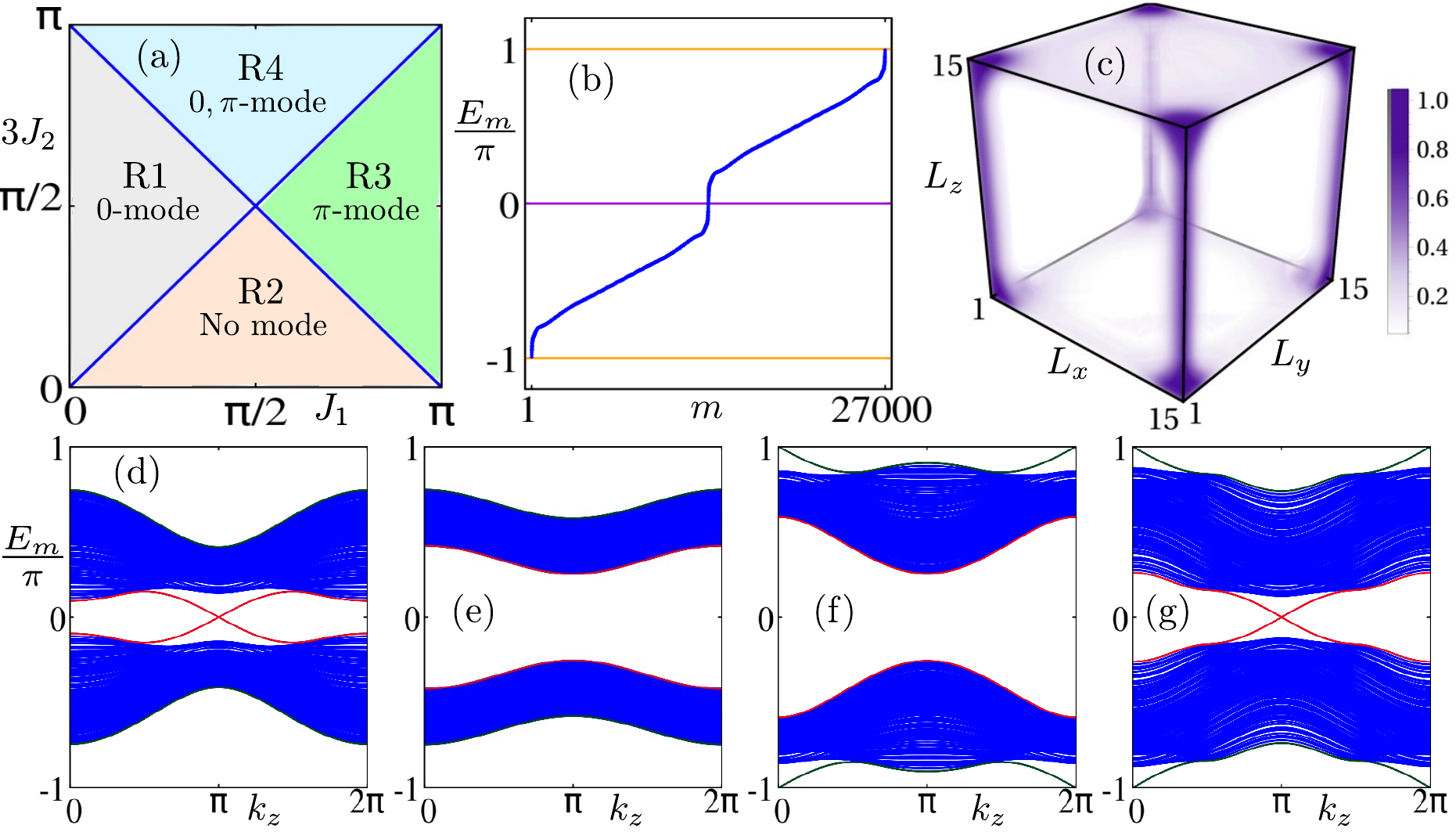}}
	\caption{
	(Color online) (a) We demonstrate the phase diagram in the $J_1-J_2$ plane for 3D FSOTI. (b) Quasi-energy spectrum, $E_m$ for finite system is shown as a function of the state index, $m$ for R4. 
	(c) The corresponding LDOS is depicted considering quasi-states with $E_m=0,~\pm \pi$ in R4. The quasi-energy spectra, $E_m$ in rod geometry along $k_z$, corresponding to R1, R2, R3, and R4 
	are shown in panels (d), (e), (f), and (g), respectively. The parameters are chosen to be the same as in Fig.~\ref{3Dfirst}.
	}
	\label{3Dsecond}
\end{figure}

In order to obtain the FFOTI phase in 3D, we set both $\alpha=0$ and $\beta=0$. The FFOTI phase is characterized by the appearance of gapless 2D surface-states. The quasi-energy spectra for a 
finite-size system with OBC along $x,~y$ and $z$-directions residing in R4, is shown in Fig.~\ref{3Dfirst}(b), as a function of the state index $m$. We depict the signature of the surface states in the LDOS corresponding to quasi-states with $E_m=0,~\pm \pi$ for R4 in Fig.~\ref{3Dfirst}(c). For a better understanding of the nature of the surface states, we employ slab geometry, by considering OBC along 
one direction (say $z$-direction), while remaining two directions obey PBC (say $x$ and $y$-direction). We depict the gapless surface states along $\Gamma-X-S-Y-\Gamma$ points, for R1, R2, R3, and R4 in Figs.~\ref{3Dfirst}(d), (e), (f), and (g), respectively. Here, $\Gamma=(0,0),~X=(\pi,0),~S=(\pi,\pi),~Y=(0,\pi)$. 
\vspace{-0.2cm}

\subsubsection*{\rm Case 2: FSOTI}
To realize the FSOTI phase, we set $\alpha$ to a non-zero value, while keeping $\beta=0$. We first explore the quasi-energy spectra under OBC in all directions as shown in Fig.~\ref{3Dsecond}(b), 
where the existence of both $0$ and $\pi$-modes in R4 are clearly visible as a function of the state index $m$. We further illustrate the LDOS corresponding to quasi-states with $E_m=0,~\pm \pi$ for R4, in Fig.~\ref{3Dsecond}(c). One can notice that the mode is populated throughout the hinge along the $z$-direction of the system. To exhibit the dispersive nature of the gapless hinge modes, we tie up to rod geometry \ie considering OBC in two directions ($x$ and $y$-direction) and PBC in the remaining direction ($z$-direction). The corresponding quasi-energy spectra for this system are shown in the rod geometry for R1, R2, R3, and R4 in Figs.~\ref{3Dsecond}(d), (e), (f), and (g), respectively. These numerical findings can be analytically understood as well from the bulk. The first Wilson-Dirac mass term proportional to $(\cos k_x -\cos k_y)$ is able to gap out the surface modes over all the three $xy$, $yz$ and $zx$ surfaces except along $x=\pm y$ for any value of $z$~\cite{Nag2020}. 

\begin{figure}[]
	\centering
	\subfigure{\includegraphics[width=0.48\textwidth]{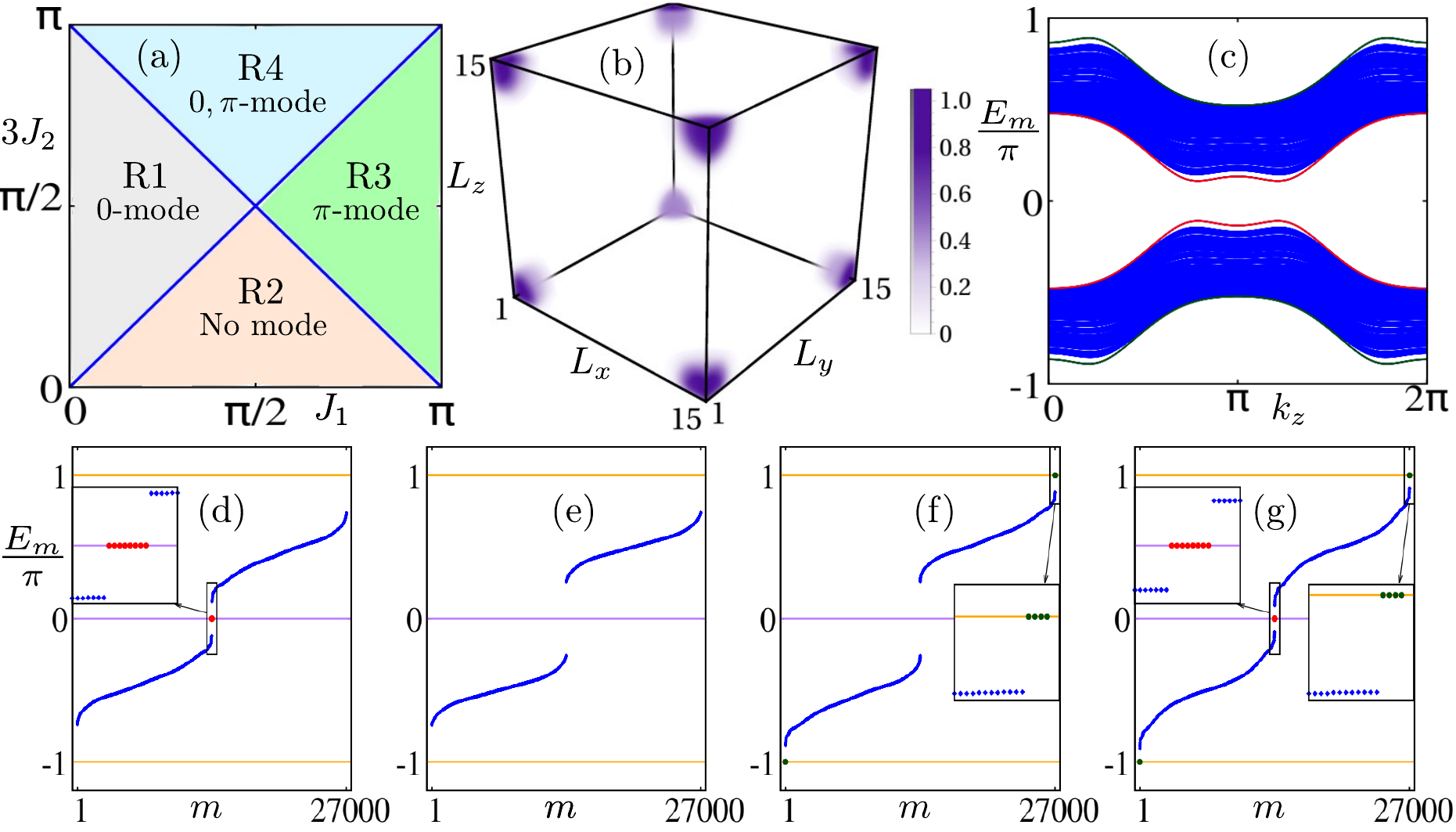}}
	\caption
	{(Color online)
	(a) We illustrate the phase diagram in the $J_1-J_2$ plane for 3D FTOTI. (b) Corner localized modes corresponding to quasi-energies $E_m=0,~\pm \pi$ is shown via LDOS for finite-size system 
in R4. (c) Quasi-energy spectrum, $E_m$ for the system in rod geometry is depicted as a function of $k_z$, manifesting the hinge modes (both $0$ and $\pm \pi$) are gapped out. The quasi-energy 
spectra $E_m$, for a finite size system and as a function of the state index $m$, corresponding to R1, R2, R3, and R4 are shown in panels (d), (e), (f), and (g), respectively. We choose the same parameter values as mentioned in Fig.~\ref{3Dfirst}.
	}
	\label{3Dthird}
\end{figure}

\subsubsection*{\rm Case 3: FTOTI}
The FTOTI phase can further be obtained when both the mass terms $\alpha$ and $\beta$ are non-zero. The corresponding phase diagram in $J_1-J_2$ plane is shown in Fig.~\ref{3Dthird}(a). In 
Fig.~\ref{3Dthird}(b), we further depict the signature of corner modes at $E_m=\pm \pi$ for R4 via the LDOS spectrum. This has been computed considering OBC in all directions. 
In the FTOTI phase, one obtains $0$ or $\pi$ mode only at the corners of the system, while both surface and hinge modes are gapped out. We show the quasi-energy spectrum for this system, employing rod geometry (OBC along $x$ and $y$-directions, PBC along $z$-direction), in Fig.~\ref{3Dthird}(c) (corresponding to R4). It is evident that, both $0$ and $\pi$ hinge modes are gapped out (see Fig.~\ref{3Dthird}(c)). Therefore, this is in contrast to the FSOTI phase as shown in Fig.~\ref{3Dsecond}(g). The second Wilson-Dirac mass term $\beta (2 \cos k_z-\cos k_x -\cos k_y)$ 
gaps out the hinge mode except at the body diagonal position $\pm z=\pm y=\pm x$ where corner modes appear~\cite{Nag2020}. The presence of dynamical corner modes can be better understood by exploring the quasi-energy spectra for a system obeying OBC in all three directions. We show the corresponding quasi-energy spectra as a function of the state index $m$, employing OBC for R1, R2, R3, and R4 in Figs.~\ref{3Dthird}(d), (e), (f), and (g), respectively. Note that, there exists $8~(4)$ quasi-states at $E_m=0,~(\pm \pi)$ for the FTOTI phase which contribute to the corner localized LDOS. 
\vspace {0.3cm}
\subsection{Periodic mass kick}
After extensive discussion on the hierarchy of FHOTI phases starting from FFOTI phases (under step drive scheme), we now proceed to analyze our findings for the mass kick protocol as given in 
Eq.~(\ref{kick1}). In the periodic mass kick formalism, the Floquet operator $U_{d \rm D}(\vect{k},T)$ can be cast in the form: 
$U_{d \rm D}(\vect{k},T)=f_{d \rm D}(\vect{k})  \mathbb{I}+i g_{d \rm D}(\vect{k})$, with

\begin{widetext}
	\begin{eqnarray}
	f_{2 \rm D}(\vect{k})&=& \cos\left(\gamma_{2 \rm D}(\vect{k})  m\right) \cos\left( \lambda_{2 \rm D}(\vect{k}) J \right) - \sin\left(\gamma_{2 \rm D}(\vect{k}) m \right) \sin\left( \lambda_{2 \rm D}(\vect{k}) J \right)  \chi_{2 \rm D}(\vect{k}) \ , \label{fk2Dkick}  \\
	g_{2 \rm D}(\vect{k})&=& - \frac{1}{\gamma_{2 \rm D}(\vect{k}) \lambda_{2 \rm D}(\vect{k})} \eta_{2 \rm D}(\vect{k}) \sin\left(\gamma_{2 \rm D}(\vect{k})  m \right) \sin\left( \lambda_{2 \rm D}(\vect{k}) J \right) -  \sin\left(\gamma_{2 \rm D}(\vect{k})  m \right) \cos \left( \lambda_{2 \rm D}(\vect{k})  J \right) \frac{h_{1, 2 \rm D}(\vect{k})}{\gamma_{2 \rm D}(\vect{k})} \non \\
	&& -  \cos\left(\gamma_{2 \rm D}(\vect{k})  m \right) \sin \left( \lambda_{2 \rm D}(\vect{k})  J \right) \frac{h_{2, 2 \rm D}(\vect{k})}{\lambda_{2 \rm D}(\vect{k})} \ , \\
	f_{3 \rm D}(\vect{k})&=& \cos\left(\gamma_{3 \rm D}(\vect{k})  m\right) \cos\left( \lambda_{3 \rm D}(\vect{k}) J \right) - \sin\left(\gamma_{3 \rm D}(\vect{k}) m \right) \sin\left( \lambda_{3 \rm D}(\vect{k}) J \right)  \chi_{3 \rm D}(\vect{k}) \ , \label{fk3Dkick}  \\
	g_{3 \rm D}(\vect{k})&=& - \frac{1}{\gamma_{3 \rm D}(\vect{k}) \lambda_{3 \rm D}(\vect{k})} \eta_{3 \rm D}(\vect{k})  \sin\left(\gamma_{3 \rm D}(\vect{k})  m \right) \sin\left( \lambda_{3 \rm D}(\vect{k})  J \right)-  \sin\left(\gamma_{3 \rm D}(\vect{k})  m \right) \cos \left( \lambda_{3 \rm D}(\vect{k})  J \right) \frac{h_{1, 3 \rm D}(\vect{k})}{\gamma_{3 \rm D}(\vect{k})} \non \\
	&&-  \cos\left(\gamma_{3 \rm D}(\vect{k}) m \right) \sin \left( \lambda_{3 \rm D}(\vect{k})  J \right) \frac{h_{2, 3 \rm D}(\vect{k})}{\lambda_{3 \rm D}(\vect{k})} \ .
	\end{eqnarray}
\vskip -0.5cm
\end{widetext}
where the mathematical symbols carry the same definitions as before. Thus, the Floquet effective Hamiltonians for 2D and 3D systems can be obtained as
\begin{eqnarray}
H_{2 \rm D, Flq}&=&- \frac{\epsilon_{2\rm D}(\vect{k})}{\sin \left[\epsilon_{2\rm D}(\vect{k}) T\right]} g_{2\rm D}(\vect{k}) \label{EffHam2Dmass} \ , \\ 
H_{3 \rm D, Flq}&=&- \frac{\epsilon_{3\rm D}(\vect{k})}{\sin \left[\epsilon_{3\rm D}(\vect{k}) T\right]} g_{3\rm D}(\vect{k}) \label{EffHam3Dmass} \ ,
\end{eqnarray}
where, $\epsilon_{2\rm D}(\vect{k})= \frac{1}{T}\cos^{-1}\left[f_{2\rm D}(\vect{k})\right] $ and $\epsilon_{3\rm D}(\vect{k})= \frac{1}{T}\cos^{-1}\left[f_{3\rm D}(\vect{k})\right]$.

\begin{figure}[]
	\centering
	\subfigure{\includegraphics[width=0.48\textwidth]{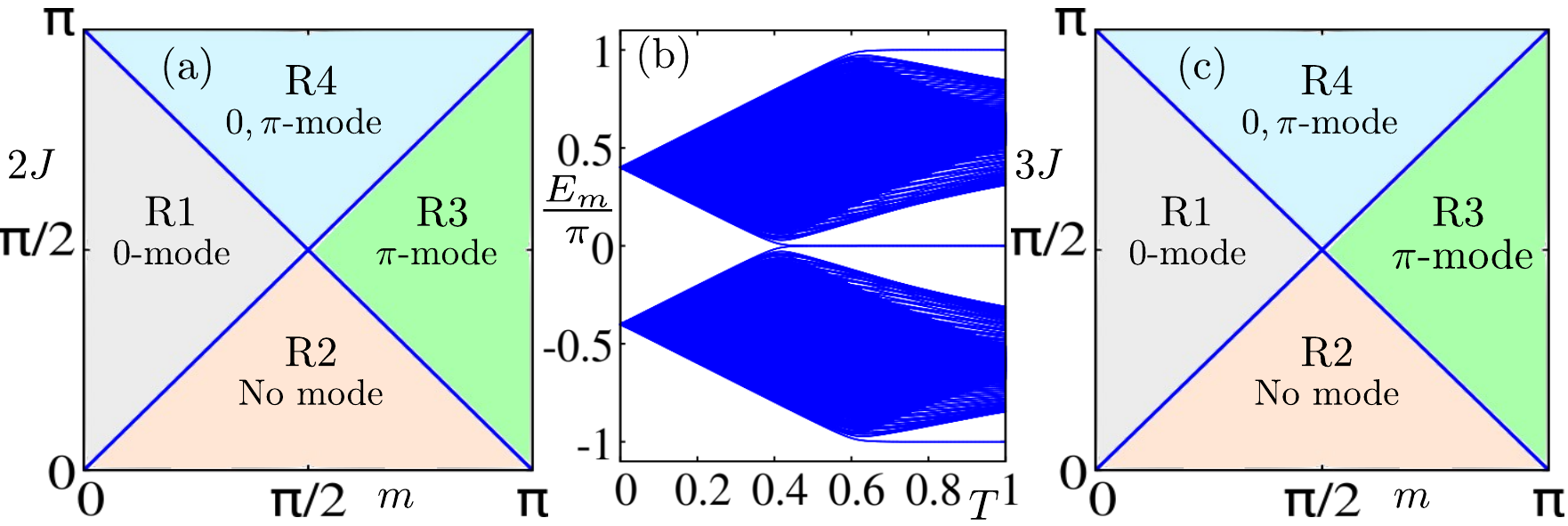}}
	\caption
	{(Color online)
	(a)	Phase diagram for the kick driving protocol is depicted in the $m-J$ plane for a 2D FFOTI/FSOTI. (b) Quasi-energy spectrum, $E_m$ is shown as a function of the driving time-period $T$. 
	This exhibits the $0,~\pm \pi$ modes. We choose $(m,J')=(0.4 \pi, 0.5 \pi)$. (c) The corresponding phase diagram for a 3D FFOTI/FSOTI/FTOTI is illustrated in the $m-J$ plane.
	}
	\label{Masskick}
\end{figure}

As discussed earlier, the gap closing conditions can be obtained using the right hand side of Eqs.(\ref{fk2Dkick}) and (\ref{fk3Dkick}) in 2D and 3D as
\begin{eqnarray}
2 \lvert J \rvert &=& \lvert m \rvert  + n \pi \  \quad {\rm :~in~ 2D} , \label{phasekick2D} \\
3 \lvert J \rvert &=& \lvert m \rvert  + n \pi \ \quad {\rm :~in~ 3D} . \label{phasekick3D}
\end{eqnarray}
Like the step drive case, the nature of phase boundary remains same for all the topological orders in every dimension. To be precise, in 2D, the phase boundary for both FFOTI and FSOTI is guided by Eq.~(\ref{phasekick2D}), whereas, in 3D, the same is controlled by Eq.~(\ref{phasekick3D}). The corresponding phase boundaries for 2D and 3D are shown in Figs.~\ref{Masskick}(a) and (c), respectively.

Here, we intend to compare our findings between mass kick and step drive protocols. The factor $\chi_{2 \rm D}(\vect{k})$, computed at the high symmetry point $\vect{k}=\vect{k}^*$,  plays important role in case of both the drives. This factor eventually allows $f_{2 \rm D}(\vect{k}^*)$ to acquire simple form in terms of a single cosine function. Once $\chi_{2 \rm D}(\vect{k}^*)$ inherits complicated mathematical form, $f_{2 \rm D}(\vect{k}^*)$ cannot be written in a compact way. More importantly, in order to observe the regular $0$-mode and anomalous $\pi$-mode together in the driven system, the condition $|\chi_{2 \rm D}(\vect{k}^*)|=1$ becomes very important. The condition $E(\vect{k}^*)=0$ ($E(\vect{k}^*)=\pm \pi$) yields the condition for obtaining $0$-mode ($\pi$-mode). 
Another crucial point to note here is that gap closing transition in the quasi-energy spectrum has to take place at the high-symmetry points \ie $\vect{k}=\vect{k}^*$. This is an essential requirement to procure the anomalous FHOTI modes. This further leads to the generic gap-closing condition, as represented by Eqs.~(\ref{cosphase2D}) and (\ref{3Dcond}), $\cos (E(\vect{k}^*))=(-1)^n$, with $n=0,1,2,3,~\cdots$. Thus the compact form of phase boundaries eventually can be portrayed as $d \lvert J_2 \rvert = \lvert J_1 \rvert  + 2n \pi$ ($d \lvert J \rvert = \lvert m \rvert  + n \pi$) for step drive (mass kick) 
with $d$ being the dimension. The above discussion is not restricted to any specific driving protocol, discussed here, rather it is applicable to a variety of driving schemes~\cite{GongPRBL2021,HuPRL2020,Huang2020}. 

For completeness, we briefly discuss the numerical results for this particular driving in 2D and 3D. To start with 2D, we obtain the FFOTI phase by setting $\alpha=0$. However, the quasi-energy spectra in the slab geometry qualitatively remains the same as of step drive case for R1, R2, R3, and R4 phase as depicted in Figs.~\ref{2Dfirst}(d), (e), (f), and (g), respectively. To procure the FSOTI, we set $\alpha$ to a non-zero value and the corresponding quasi-energy spectra encompassing the corner modes (at $E_{m}=0, \pm\pi$) for a finite-size system turns out to be similar to that shown in 
Figs.~\ref{2Dsecond}(d), (e), (f), and (g) for the sectors R1, R2, R3, and R4, respectively.

In 3D, the FFOTI emerges when both $\alpha$ and $\beta$ are zero. The surface states can be found out by realizing the corresponding lattice model in the slab geometry (see Figs.~\ref{3Dfirst}(d), (e), (f), and (g)). The qualitative nature of these surface states remain the same as obtained implementing the step drive protocol. Furthermore, the system manifests FSOTI hosting gapless hinge modes when 
$\alpha \neq 0$ but $\beta=0$. The corresponding signature is highlighted in the quasi-energy spectrum that has been calculated using rod geometry (see Figs.~\ref{3Dsecond}~(d), (e), (f), and (g)). 
To obtain the FTOTI phase, both $\alpha$ and $\beta$ are set to a non-zero value. The footprints of the corner localized modes (at $E_{m}=0, \pm\pi$) in the FTOTI phase is found out in the quasi-energy spectrum for a finite-size system (see Figs.~\ref{3Dthird}(d), (e), (f), and (g)).

One intriguing difference regarding topological phase boundary equations between step drive and periodic kick is absence of the time-period $T$ in the latter case. Importantly, for the step drive, the time period $T$ is coupled to both the driving parameters $J'_{1,2}$ such that $J_{1,2}$ becomes the effective parameter to characterize the dynamical system. On the other hand, for periodic kick, there exists two effective driving parameters $J=J'T$ and $m$ where only one of them is renormalized by $T$ while the other remains unaltered. This, in fact, allows us to seek for a frequency driven topological phase transition in the system (see Fig.~\ref{Masskick}(b)). We begin with a parameter set: $(m,J')=(0.4 \pi,0.5 \pi)$, such that we belong to R2 with no modes. Afterwards, we decrease~(increase) the frequency~(time period) of the drive, while keeping both $m$ and $J'$ fixed. Thus, we move towards R1, where the $0$-mode appears, and then to R4, where both the $0$ and $\pi$-mode emerge. Therefore, the topological  phase transition, mediated by the driving parameters, is a common feature for step driving~\cite{GongPRBL2021,HuPRL2020,Huang2020} while the kick protocol can further give rise to the frequency driven topological phase transition~\cite{nag2021anomalous}. 
\vspace {-0.4cm}
\section{Topological Characterization}\label{Sec:IV}
The topological protection of corner modes in 2D SOTI can be traced down by the position resolved tangent polarization~\cite{benalcazarprb2017}. Owing to the fact that quadrupole can be thought of in terms of dipole  pumping, the fractional corner charge in 2D can be traced back to tangential polarization, defined in the semi-infinite geometry with OBC ($L_x$ number of sites) along one of the direction (say $x$). Motivated by the above analogy in static system, we below examine the tangential polarization for the driven case to characterize FSOTI~\cite{Franca2018}. To compute the following, we first construct the Wilson loop operator~\cite{benalcazarprb2017} in the slab geometry (considering PBC along $y$-direction and OBC along $x$-direction) as ${\mathcal W}_y=F_{y,k_y + (N_y -1) \Delta k_y } \cdots F_{y,k_y + \Delta k_y } F_{y,k_y} $ with $ \left[F_{y,k_y}\right]_{mn}=\langle \psi_{n, k_y + \Delta k_y} | \psi_{m,k_y} \rangle$, where $\Delta k_y= 2\pi /N_y$ ($N_y$ being the number of discrete points considered inside the Brillouin zone (BZ) along $k_y$), $|\psi_{m,k_y} \rangle $ is the $m^{\rm th}$ occupied quasi-energy state of the Floquet operator $U_{2 \rm D}(k_y,T)$, and $k_y$ denotes the base point from where we start to construct the Wilson loop operator. 
\begin{figure}[H]
	\centering
	\subfigure{\includegraphics[width=0.5\textwidth]{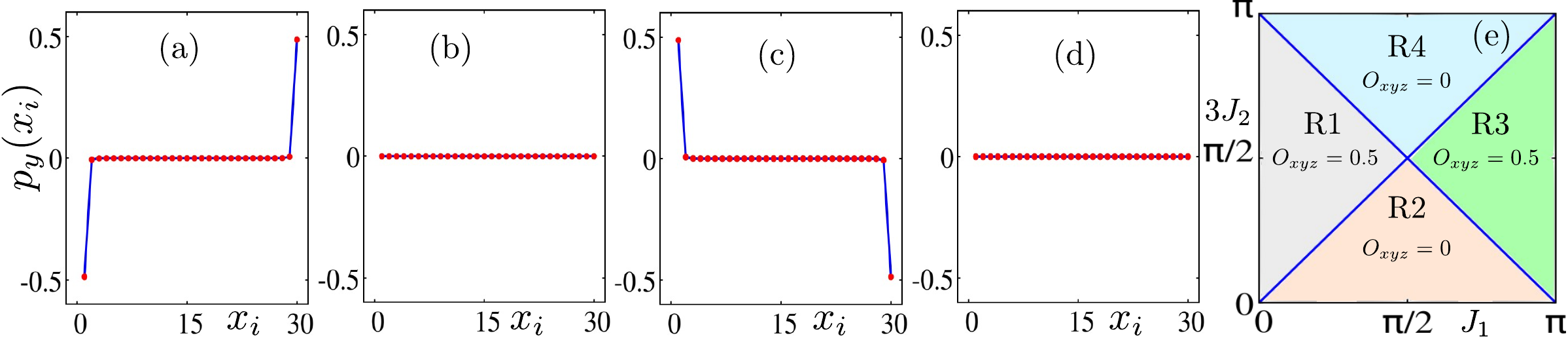}}
	\caption
	{(Color online)
		Tangential polarisation $p_y(x_i)$ is demonstrated for FSOTI for R1, R2, R3, and R4 in panels (a), (b), (c), and (d) respectively, choosing the same parameters used as in Fig.~\ref{2Dsecond}. (e) Octupolar moment, $O_{xyz}$ is schematically shown for a 3D  
                  FTOTI in the $J_1-J_2$ plane. $O_{xyz}=0.5$ in phases R1 and R3, whereas $O_{xyz}=0.0$  
                  in both phase R2 and R4.
	}
	\label{tanPol}
\end{figure}
\noindent
Then, the $2N \times 2N$ dimensional Wannier Hamiltonian ${\mathcal H}_{{\mathcal W}_y}$ can be written as ${\mathcal H}_{{\mathcal W}_y}= -i \ln {\mathcal W}_y$, whose eigenvalues are of the form $2\pi \nu_y$. Here, $\nu_y \equiv {\rm mod} (\nu_y,1)$ is the Wannier center. The position dependent tangential polarization $p_y(x_i)$ is defined as
\vskip -0.3cm
\begin{align}
p_y(x_i)=\frac{1}{2 \pi  N_y} \sum_{j,k_y,\beta} \left|  \sum_{\alpha} \left[\nu_{j,k_y}\right]_{\alpha} 
\left[ \psi_{\alpha,k_y} \right]_{\beta,x_i} \right|^2 \nu_{j,y} \ ,
\label{edpol}
\end{align}
\vskip -0.3cm
\noindent
where, $\left[\nu_{j,k_y}\right]_{\alpha}$ refers to the $\alpha^{\rm th}$ component of the $j^{\rm th}$ eigenstate $\ket{\nu_{j,k_y}}$ of the Wannier Hamiltonian ${\mathcal H}_{{\mathcal W}_y}$, corresponding to the Wannier center $\nu_{j,y}$. Although the Wanner center $\nu_{j,y}$ is independent of the base point $k_y$, however the eigenstates of the Wannier Hamiltonian do 
depend on the base point. Here, $\left[ \psi_{\alpha,k_y} \right]_{\beta,x_i}$ is the 
$\left(\beta,x_i\right)^{\rm th}$ component of the $\alpha^{\rm th}$ occupied eigenstate 
$\ket{\psi_{\alpha,k_y}}$ corresponding to the Floquet operator $U_{2 \rm D}(k_y,T)$. The index $\beta$ denotes the combined number of spin and pseudo-spin degrees of freedom at a lattice site, which is 4 in the present case. 

We depict the behavior of $p_y (x_i)$ as a function of the lattice site $x_i$ in Figs.~\ref{tanPol}(a), (b), (c), and (d) corresponding to the phase R1, R2, R3, and R4 (see earlier text and Fig.~\ref{2Dsecond}(a)), respectively. In the R1 and R3 phase, the system exhibits corner modes at quasi-energy $E_m=0$ or $\pm \pi$ (see Figs.~\ref{2Dsecond}(d) and (f)). In these cases, the tangential polarisation is able to portray the topological nature of corner modes. In particular, the edge-polarization, $p_y^{\rm edge}=\sum_{x_i=1}^{L_x/2} p_y(x_i) $ exhibits a quantized value of $0.5$, which is the signature of second-order topological phase~\cite{benalcazarprb2017}; $L_x$ being the number of lattice sites along the $x$-direction. In contrast, $p_y^{\rm edge}$ remains at $0$ for both the trivial phase R2 and the non-trivial phase R4 (hosting both 0 and $\pm\pi$ modes) and hence unable to conclude any apprehensible distinction between them~\cite{Huang2020}. The dynamical quadrupolar motion introduced in Ref.~\cite{Huang2020}, however, can unambiguously reveal the presence of corner mode at $0$ or $\pi$-gap. But to evaluate the said invariant, one needs to own mirror symmetry present in the system, which is not the case here. Therefore, calculation of a topological indicator for a mirror symmetry broken model system remains an open question.

We here emphasize that the breaking of mirror symmetries in $h_{2,{d \rm D}}(\vect{k})$ severely affect the quadrupole and octupole motion that otherwise can distinguish regular 
$0$-mode from anomalous $\pi$-mode by exhibiting gapless crossing with time and accurately capturing the bulk boundary correspondence for the driven systems~\cite{Huang2020,ghosh22a}. The above two quantities serve as the legitimate bulk invariants for the FHOTI phases following the construction of dynamical nested Wilson loop. We cannot resort to these bulk invariants for their gapped profile due to the absence of mirror symmetry in our case. Therefore, we continue with the edge-polarization (Eq.~(\ref{edpol})) which turns out to be an inappropriate bulk invariant for dynamics and thus the bulk boundary correspondence cannot be fully captured for all the dynamical phases discussed here.

For a FTOTI in 3D, considering PBC in the real space geometry, one can calculate the octupole moment, defined as~\cite{Kang2019PRB}
\begin{eqnarray}
O_{xyz}&=& {\rm Re}\left[ -\frac{i}{2 \pi} {\rm Tr}\left( \ln \Big(\Psi_{0}^\dagger \exp\Big[2\pi i \sum_r  {\hat o}_{xyz} (r)\Big] \Psi_{0} \Big)\right) \right]. \non \\
\label{om_xy}
\end{eqnarray}
\noindent
where, $\Psi_{0}$ is the many-body ground state which we obtain by columnwise arranging the quasi-energy states of the Floquet operator $U_{3 \rm D}(T)$ according to their quasi-energy 
$-\Omega/2 \le E_m \le 0$: $\Psi_{0}=\sum_{m \in E_m \le 0} |\phi_m \rangle \langle \phi_m|$~\cite{Nag19,Nag2020,Ghosh2020}. Also, ${\hat o}_{xyz}={\hat n}(r) x y z/L^3$ with ${\hat n}(r)$ being the number operator at $r=(x,y,z)$.

Note that, for the phase R1 and R3, $O_{xyz}$ exhibits a quantized value of $0.5$, where corner modes appear separately at quasi-energy $E_m=0$ and $\pm \pi$, respectively~(see Fig.~\ref{3Dthird}). Although $O_{xyz}$ cannot discriminate between the phase R2 and R4, manifesting a congruent value of $0$. We depict $O_{xyz}$ using a schematic representation in Fig.~\ref{tanPol}(e). 
We note that the octupole moment can be considered as the appropriate bulk invariant for static system. The anomalous mode cannot be appropriately characterized by the same. 
It might be possible that $0$- and $\pi$-mode interfere destructively yielding $0$ (mod $1$) as a measure of the invariant. As discussed above, the Floquet operator again turns out to be insufficient for 
the proper dynamical characterization. Therefore, octupole moment (Eq.~(\ref{om_xy})) as well cannot refer to the accurate bulk boundary correspondence for driven system similar to the 
edge-polarization (Eq.~(\ref{edpol})).

\section{Discussions and Outlook}\label{Sec:V}
\begin{figure}[]
	\centering
	\subfigure{\includegraphics[width=0.48\textwidth]{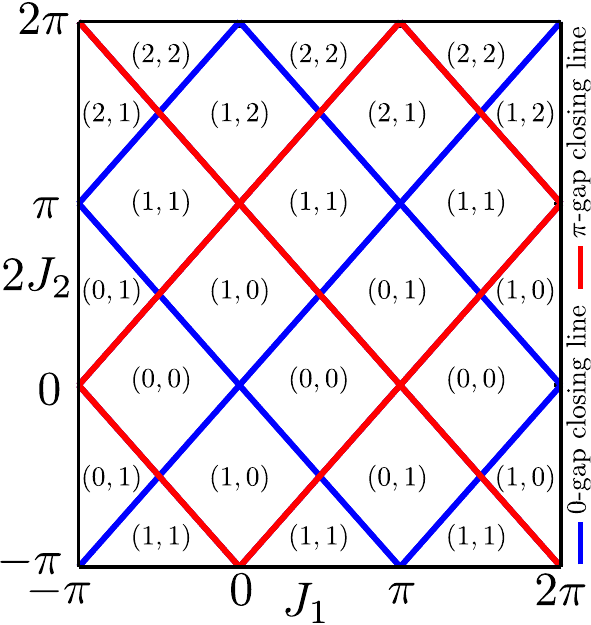}}
	\caption
	{(Color online)
		Extended phase diagram is illustrated in the $J_1-J_2$ plane in case of a 2D FFOTI/FSOTI. Here, $(i,j)$ represent the number of $0, \pi$-modes present at that parameter regime.
	}
	\label{Extd}
\end{figure}

{\bf{\textit{Extended phase diagram:}}} First our aim is to generalize the topological phase diagram as displayed in Fig.~\ref{2Dsecond}(a). In order to contemplate the extended phase diagram, we extend the scale of $(J_1,J_2)$ so that one can obtain various phases $|J_2|=|J_1|/2 + n \pi$ with different values of $n$. We know that there exist $4$ ($2$) regular $0$ (anomalous $\pi$, residing at $E_m=+\pi$ or $-\pi$)-mode in the FSOTI phase. By changing the parameter set $(J_1,J_2)$, one can in principle reach to several other FSOTI phases where the number of $0$ and $\pi$ modes can be varied. This is extensively demonstrated in Fig.~\ref{Extd}. The $0$ ($\pi$)-gap closings are indicated by the blue (red) lines. As a result, one can clearly finds that the dynamical topological phases, separated by blue (red) lines, are hosting the identical number of $\pi$ ($0$)-mode. The blue (red) lines are associated with $n=0,2,4,\cdots$ ($n=1,3,5,\cdots$) provided $J_1, J_2$ being positive. 
More interestingly, this phase diagram is invariant with the time period $T$ as the effective dynamics controlling parameters $(J_1,J_2)$ are both renormalized by the time period $T$. One can hence think of this extended phase diagram is equally valid for the other step drive protocols~\cite{GongPRBL2021,HuPRL2020,Huang2020}. However, the above discussion is applicable for all the other phase 
diagrams for step drive scheme presented in this work. Therefore, in future it would be an interesting open question to characterize these phase with appropriate dynamical topological invariant.

\vspace{0.1cm}
{\bf{\textit{High frequency approximation:}}}
Having discussed and analyzed our findings at a given frequency, we now present the high frequency effective Hamiltonians for both the driving protocols. In the high-frequency limit \ie $T \rightarrow 0$, the effective Hamiltonians for 2D and 3D in case of step drive and the periodic mass kick boil down to the following form~\cite{mikami16}:
\small
\begin{eqnarray}
H_{2 \rm D, step}^{\rm eff} (\vect{k})&=& \frac{J_1'}{2} h_{1, 2 \rm D}(\vect{k}) + \frac{J_2'}{2} h_{2, 2 \rm D}(\vect{k}) + \frac{J_1' J_2' T}{4} \eta_{2 \rm D}(\vect{k}) \ , \qquad \label{2DstepHigh} \\
H_{3 \rm D, step}^{\rm eff}(\vect{k}) &=& \frac{J_1'}{2} h_{1, 3 \rm D}(\vect{k}) + \frac{J_2'}{2} h_{2, 3 \rm D}(\vect{k}) + \frac{J_1' J_2' T}{4} \eta_{3 \rm D}(\vect{k}) \ ,\label{3DstepHigh} \\
H_{2 \rm D, kick}^{\rm eff}(\vect{k}) &=& J' h_{2, 2 \rm D}(\vect{k}) + \frac{m}{T} h_{1, 2 \rm D}(\vect{k}) + m J' \eta_{2 \rm D}(\vect{k}) \ ,\label{2DkickHigh} \\
H_{3 \rm D, kick}^{\rm eff} (\vect{k})&=& J' h_{2, 3 \rm D}(\vect{k}) + \frac{m}{T} h_{1, 3 \rm D}(\vect{k}) + m J' \eta_{3 \rm D}(\vect{k}) \ . \label{3DkickHigh}
\end{eqnarray}

\normalsize
For the mass kick protocol, we also approximate $m \rightarrow 0$ such that $m/T$ becomes finite. One can observe that the static Hamiltonians $H_{2 \rm D}^{\rm Static}(\vect{k})$ and $H_{3 \rm D}^{\rm Static}(\vect{k})$ (Eqs.(\ref{2DStatic}) and (\ref{3DStatic})) resemble the high-frequency effective Hamiltonians $H_{2 \rm D, step}^{\rm eff} (\vect{k})$ and $H_{3 \rm D, step}^{\rm eff} (\vect{k})$ (Eqs.(\ref{2DstepHigh}) and (\ref{3DstepHigh})), except the apprearance of the additional term associated with $\eta_{d \rm D}(\vect{k})$ in the latter case. In the high frequency limit, the effective Hamiltonians are quasi-static in the sense that no long range hopping has been incorporated with the inclusion of this additional term proportional to 
$\eta_{d \rm D}(\vect{k})$. This might allow us to work with the static definition of topological invariant that are equally trustworthy to characterize the dynamical phases such as Floquet quadrupole moment \cite{Nag19,Nag2020,Ghosh2020,ghosh2020floquet,ghosh2020floquet2,ghosh2021ladder}. We note that the $0$ mode is expected to survive even in presence of this extra factor as the effective Hamiltonians preserve particle-hole symmetry: ${\mathcal P} H_{d \rm D,\zeta }^{\rm eff} (\vect{k}) {\mathcal P}^{-1}=- H_{d\rm D,\zeta}^{\rm eff} (-\vect{k})$ with $\zeta$= step,~kick \cite{Nag19,broy20}. Interestingly, with decreasing frequency, different Floquet zones can come closer that might results in the anomalous $\pi$-mode.  However, the above statement is not equally applicable to different kinds of driving. It might be useful in future to analyze the dynamical topological phases, with appropriate topological characterization, in the intermediate frequency range where one can consider 
${\mathcal O}(T^n)$ for $n>1$ terms. 
    
{\bf{\textit{Other possible driving schemes to engineer FHOTI:}}} We know that one can architect the FFOTI phase by employing laser driving as extensively discussed for 2D lattice systems~\cite{oka09photovoltaic,nag17,lindner11floquet,Usaj2014,Mohan2014,Kundu2016}. Very recently, the emergence of FSOTI phases in 2D has been introduced using laser driving~\cite{Ghosh2020}. On the other hand, periodic driving through phonon mediated spin-orbit coupling is shown to exhibit FSOTI phase when the underlying static system has reflection symmetry~\cite{chaudhary20}. The manipulation of TRS using Floquet-Zeeman term becomes instrumental to experience Floquet second order topological superconductor phase~\cite{plekhanov19}.  The role of mirror symmetry is also emphasized in order to achieve the FSOTI phase~\cite{Martin2019}. Having discussed a handful of examples including $C_4$ symmetry breaking perturbations~\cite{Nag19,schindler2018}, we can comment that only a few of the available dynamical protocols are able to produce the anomalous $\pi$ mode in the FHOTI phases. In this context, the gap-structure of the quasi-energy spectrum at the high symmetry point might become very important as we discuss previously. Our approach is successfully able to portray the series of FHOTI phases, FFOTI $\to$ FSOTI $\to$ FTOTI, while incorporating different symmetry breaking terms only in one of steps (mid-kick) Hamiltonian for step (kick)-protocol. In the present case, Floquet operator satisfies anti-unitary symmetry ${\mathcal P} U_{d \rm D}(\vect{k},T){\mathcal P}^{-1}= U_{d \rm D}(-\vect{k},T)$ referring to the fact that the $0$ and $\pi$-modes in FFOTI and FHOTI phases protected by the particle-hole symmetry~\cite{rroy17}. It remains an open question so far that how one can engineer the hierarchy of FHOTI phases (hosting dynamical $\pi$-modes) in 3D employing other drive protocols (\eg laser driving). 

We now compare our findings with another related work where systematic generation of regular FHOTI phases with zero quasi-energy modes has been demonstrated~\cite{Nag2020}. The periodic kick with high-frequency, compared to the band width of the system, in the Wilson-Dirac mass term there leads to the zero quasi-energy Floquet hinge and corner modes. 
A quantitative investigation, adopted from a specific case with the kick in the second order Wilson-Dirac mass term $V_1=\sqrt{3} \Delta_1(\cos k_x - \cos k_y)$ ($\Delta_1<t_0$), suggests that to observe regular FHOTI phase with zero quasi-energy, the time period $T$ has to be small compared to the band width of the static system with $t_0$ and $m$ being hopping amplitude and first order mass term, respectively. Interestingly, the anomalous phase (with $\pi$-mode) cannot be embedded in the high-frequency phase diagram following the above drive protocol~\cite{Nag2020}.
This is in stark contrast to our present case with step drive protocol where all four dynamic phases R1, R2, R3, and R4 have emerged irrespective of the frequency regime.

Having compared the various Floquet engineering of HOT phases, we now highlight the structural tunability of Floquet hinge modes. It has been shown that the SOTI phase in 
3D can host connected hinge modes, preserved by a nonsymmorphic space-time symmetry namely, time glide symmetry, under harmonic drive~\cite{YangPRL2019}. To be precise, the hinge modes are localized at the intersections of the two sets of surfaces that are related by the reflection part of the time-glide symmetry. The harmonic drive induces opposite masses in these sets of surfaces leading to connected hinge at their intersections along $\tilde {x}$-, $\tilde {z}$-, and $\tilde {y}$-directions provided the crystal cuts are not along the principle axes. We do not encounter such hinge modes in our case where four disconnected hinge modes appear only along $z$-direction as $xz$ and $yz$-surfaces are gapped out with opposite mass terms. Therefore, the SOTI is manufactured out of the FOTI by gapping out the surface modes except at the hinges in 3D due to the Wilson Dirac mass term $\left(\cos k_x - \cos k_y \right) \mu_x \sigma_y$~\cite{schindler2018,Roy2019,Nag2020,ghosh2020floquet2}.  Importantly, the SOTI phase is preserved by $C_4 {\mathcal T}$ symmetry. The above mechanism also holds for the FSOTI phase as long as lattice termination remains compatible with the $C_4$ symmetry, respecting the principle axes. The Floquet dynamics can be simply understood by the fact that the FFOTI is elevated to FSOTI followed by FTOTI while the HOT mass terms are appropriately incorporated in  $h_{2,{\rm 3D}}(\vect{k})$. Therefore, one can in principle engineer disconnected and connected hinge modes by suitably choosing the Wilson-Dirac mass and/or Floquet driving and/or 
lattice configuration.
    
{\bf{\textit {Effect of disorder:}}} In recent years, the disorder mediated HOTI phases have attracted significant attention~\cite{araki19,ybyang21,jhwang21,yshu21}. In particular, random on-site disorder in 
2D system, with chiral symmetry, can induce a quadrupolar topological insulating phase with interesting localization properties~\cite{ybyang21}. On the the hand, for driven systems, the concept of Floquet Anderson insulator is introduced where topologically protected non-equilibrium transport phenomenon is observed~\cite{titum16,torres19}. In this context, the effect of weak disorder has been investigated 
in Floquet quasi-static second order topological superconductor phase~\cite{ghosh2020floquet2}. This is an emerging field of research where strong disorder effects can lead to significantly different phenomena in the the context of FHOTI phases~\cite{jsong12,modak20}. The appropriate definition of topological invariant would become very important to characterize these phases which remains 
an open field of research.  

\section{Summary and Conclusion}\label{Sec:VI}
\vspace{-0.1cm}
To summarize, in this article we consider simple periodic drive protocols and illustrate the systematic generation of series of FHOTI phases including the FFOTI phase, both in 2D and 3D. To begin with, 
we consider the two-step drive protocol where the first half is comprised of on-site terms only and the second half contains all the off-diagonal hopping terms. With such an admixture of tight binding Hamiltonians, we first show the appearence of FFOTI phase in absence of the discrete symmetry breaking Wilson-Dirac mass terms. Upon systematic inclusion of these terms in the second step Hamiltonian, we obtain the FSOTI followed by FTOTI phases in 3D. In the process, we also exemplify the 2D case where one can reach upto corner modes in FSOTI phase. Most interestingly, we find regular quasi-static $0$-mode, similar to the static case, as well as anomalous (dynamical) $\pi$-mode, unlike to the static case, in all the above phases. We analytically study the evolution operator and the corresponding topological phase diagram, consisting of $0$, $\pi$, $0$-$\pi$-modes in different parameter regime. The latter is successfully explained by the appropriate gap closing conditions from the evolution operator. We support our analytical findings by numerical computation for finite-size system with appropriate boundary conditions. We further continue with another driving scheme namely, mass kick to recheck the robustness of the above findings. The qualitatively identical phase diagram, obtained under the above driving scheme, can be attributed to the special structure of the identity term in the evolution operator at the high symmetry points. Most interestingly, the number of $0$ and $\pi$-modes can vary in the step drive case while frequency driven topological phase transition is witnessed 
for the mass kick protocol. We employ tangential polarization (Floquet octupolar moment) to topologically characterize the FSOTI (FTOTI) in 2D (3D) with either $0$ or $\pm \pi$ corner modes. 
However, our topological inviariants are incapable of characterizing the dynamical phase hosting regular $0$ and anomalous $\pi$-mode simultaneously.

In recent times, significant experimental advancement has been put together in case of HOTI based on solid state materials~\cite{schindler2018higher,Experiment3DHOTI.VanDerWaals},
accoustic systems~\cite{xue2019acoustic,ni2019observation,Experiment3DHOTI.aSonicCrystals,serra2018observation}, classical electrical circuits~\cite{imhof2018topolectrical} etc. 
Although, the experimental observation of FHOTI phases in 2D and 3D is still in its infancy~\cite{WeiweiZhu2020}. Neverthess, given the experimental progress in this research field, we believe that
our theoretical model and driving protocols to generate the series of FHOTI phases (anchoring $0$ and $\pi$ modes) is timely and may be possible to realize in future experiments. However, the exact description of experimental techniques and prediction of candidate material are not the subjects of our present manuscript.

\vspace {0.5cm}
\subsection*{Acknowledgments}
 A.K.G. and A.S. acknowledge SAMKHYA: High-Performance Computing Facility provided by Institute of Physics, Bhubaneswar, for numerical computations. T.N. acknowledges Bitan Roy for useful discussions. 


\bibliography{bibfile}{}

\end{document}